\documentclass[aps,twocolumn,pra,superscriptaddress,showpacs,tightenlines]{revtex4-1}
\usepackage [latin1]{inputenc}
\usepackage{amsmath}
\usepackage{graphicx}
\usepackage{color}
\usepackage{amsfonts}
\usepackage{txfonts}
\usepackage[colorlinks,citecolor=blue]{hyperref}
\begin{document}

\title{Significant enhancement in refrigeration and entanglement in auxiliary-cavity-assisted optomechanical systems}

\author{Deng-Gao Lai}
\affiliation{Theoretical Quantum Physics Laboratory, RIKEN Cluster for Pioneering Research, Wako-shi, Saitama 351-0198, Japan}

\author{Wei Qin}
\affiliation{Theoretical Quantum Physics Laboratory, RIKEN Cluster for Pioneering Research, Wako-shi, Saitama 351-0198, Japan}

\author{Bang-Pin Hou}
\affiliation{College of Physics and Electronic Engineering, Institute of Solid State Physics, Sichuan Normal University, Chengdu 610068, P. R. China}

\author{Adam Miranowicz}
\affiliation{Theoretical Quantum Physics Laboratory, RIKEN Cluster for Pioneering Research, Wako-shi, Saitama 351-0198, Japan}
\affiliation{Institute of Spintronics and Quantum Information, Faculty of Physics, Adam Mickiewicz University, 61-614 Pozna\'{n}, Poland}

\author{Franco Nori}
\affiliation{Theoretical Quantum Physics Laboratory, RIKEN Cluster for Pioneering Research, Wako-shi, Saitama 351-0198, Japan}
\affiliation{RIKEN Center for Quantum Computing (RQC), 2-1 Hirosawa, Wako-shi, Saitama 351-0198, Japan}
\affiliation{Physics Department, The University of Michigan, Ann Arbor, Michigan 48109-1040, USA}

\begin{abstract}
We propose how to achieve significantly enhanced quantum refrigeration and entanglement by coupling a pumped auxiliary cavity to an optomechanical cavity. We obtain both analytical and numerical results, and find optimal-refrigeration and -entanglement conditions under the auxiliary-cavity-assisted (ACA) mechanism. Our method leads to a \emph{giant amplification} in the net refrigeration rate, and reveals that the ACA entanglement has a much \emph{stronger noise-tolerant ability} in comparison with the unassisted case. By appropriately designing the ACA mechanism, an effective mechanical susceptibility can be well adjusted, and a \emph{genuine} tripartite entanglement of cooling-cavity photons, auxiliary-cavity photons, and phonons could be generated. Specifically, we show that both optomechanical refrigeration and entanglement can be greatly enhanced for the blue-detuned driving of the auxiliary cavity but suppressed for the red-detuned case. Our work paves a way towards further quantum control of macroscopic mechanical systems and the enhancement and protection of fragile quantum resources.
\end{abstract}

\maketitle

\section{Introduction\label{sec1}}

Exploring radiation-pressure interactions between light and mechanical motion in cavity optomechanics~\cite{Kippenberg2008Science,Meystre2013AP,Aspelmeyer2014RMP,Bowen2015book} has lead to an impressive development of efficient methods for generating and controlling photon blockade~\cite{Rabl2011PRL,Nunnenkamp2011,Liao2012PRA,Liao2013PRA,Wang2015PRA1,Huang2018PRL,Li2019PR,Zou2019PRA,Liao2020PRA}, optomechanically induced
transparency~\cite{Agarwal2010PRA,Weis2010Science,Safavi-Naeini2011Nature,Wang2014PRA,Hou2015PRA,Lai2020PRA1}, dynamical Casimir effect~\cite{Cirio2017PRL,Stefano2019PRL,Qin2019PRA,Wang2019PRA}, and nonreciprocal excitation transport~\cite{Malz2018PRL,Shen2016NP,Shen2018NC,Fang2017NP,Xu2019Nature,Mathew2018arXiv,Yang2020NC,Xu2016Nature,SanavioPRB2020}. In particular, optomechanical cooling~\cite{Wilson-Rae2007PRL,Marquardt2007PRL,Genes2008PRA} and entanglement~\cite{Vitali2007PRL,Vitali2007JPA,Genes2008NJP,Mancini2002PRL,Paternostro2007PRL,Riedinger2018Nature,Ockeloen-Korppi2018Nature,Qin2019npj} studied here are, respectively, a prerequisite for observing and manipulating quantum mechanical effects and a key element in quantum information processing.

So far, several cooling mechanisms based on optomechanical systems, such as resolved-sideband cooling~\cite{Wilson-Rae2007PRL,Marquardt2007PRL} and feedback-aided cooling~\cite{Mancini1998PRL,Genes2008PRA,Steixner2005PRA,Bushev2006PRL,Rossi2017PRL,Rossi2018Nature,Conangla2019PRL,Tebbenjohanns2019PRL,Sommer2019PRL,Guo2019PRL,Sommer2020PRR}, have been proposed to cool mechanical resonators to their quantum ground states. To further develop cooling performance, various new cooling schemes have been proposed, such as those based on quantum interference~\cite{Wang2011PRL,Li2011PRB,Yan2016PRA}, parity-time symmetric~\cite{Liuyu2017PRA}, modulated pulses~\cite{Liao2011PRA,Machnes2012PRL}, domino effect~\cite{Lai2018PRA,Lai2021PRA}, strong couplings~\cite{Liu2013PRL,Liu2014PRA}, and nonreciprocity~\cite{Lai2020PRARC,Xu2019Nature}.
Particularly, cooling of mechanical resonators has also been simultaneously achieved in optical~\cite{Chan2011Nature,Teufel2011Nature,Clarkl2017Nature,MXu2020PRL,Qiu2020PRL} and microwave~\cite{Grajcar2008PRB,Zhang2009PRA,Liberato2011PRA,Xue2007PRB,You2008PRL,Nori2008NP,Xiang2013RMP} platforms. These theoretical and experimental advances enable the generation of nonclassical mechanical states and the quantum manipulation of macroscopic mechanical systems.

In parallel, optomechanical interfaces also provide a powerful tool for achieving quantum entanglement between, e.g., a cavity-field mode and a mechanical mode, two cavity-field modes, and two mechanical resonators~\cite{Vitali2007PRL,Vitali2007JPA,Genes2008NJP,Mancini2002PRL,Paternostro2007PRL,Riedinger2018Nature,Ockeloen-Korppi2018Nature}.
However, this generated entanglement is often limited by the stability conditions of the systems~\cite{Vitali2007PRL,Vitali2007JPA,Genes2008NJP} and the amplification effect in the unstable regime~\cite{Hofer2011PRA,Vanner2011PNAS}. In particular, environmental thermal noises can destroy fragile quantum entanglement in practical devices.
To generate highly pure quantum entanglement, reservoir engineering techniques~\cite{Wang2013PRL,Chen2014PRA,Wang2015PRA,Woolley2014PRA,Yang2015PRA,Li2015NJP,LiaoCG2018PRA}, quantum interference effect~\cite{Tian2013PRL,Genes2011PRA,Guo2014PRA,Liu2015PRA,Gu2013PRA,Feng2017PRA}, time modulation of the driving laser~\cite{Mari2009PRL,Mari2012NJP,ZLi2015PRA,Wangm2016PRA}, photon counting techniques~\cite{Ho2018PRL}, and Sagnac effect~\cite{Jiao2020PRL}, have been proposed based on cavity optomechanical systems. Despite such achievements, the enhancement of both optomechanical cooling and entanglement, and the protection of fragile quantum correlations in practical devices still need further studies.

\begin{figure*}[tbp]
\center
\includegraphics[width=1 \textwidth]{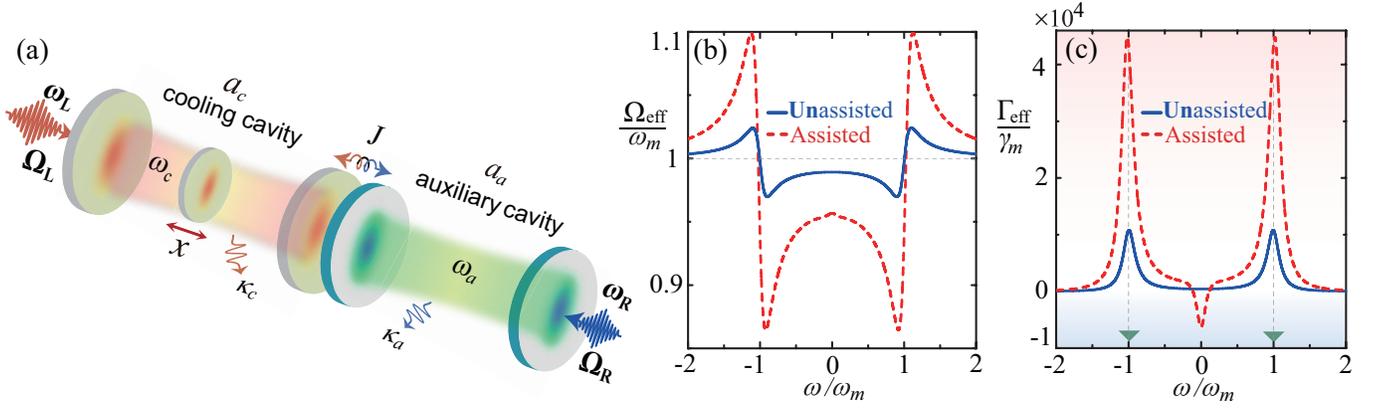}
\caption{(a) Schematics of the optomechanical system. A cooling cavity $a_{c}$ with resonance frequency $\omega_{c}$ is, respectively, coupled to a mechanical resonator with the position operator $x$, via radiation-pressure coupling with strength $g$, and to an auxiliary cavity through the tunneling coupling with strength $J$. A monochromatic laser with frequency $\omega_{L}$ ($\omega_{R}$) and field amplitude $\Omega_{L}$ ($\Omega_{R}$) is introduced to drive the cooling (auxiliary) cavity. (b) Effective mechanical frequency $\Omega_{\text{eff}}(\omega)$ [see Eq.~(\ref{effective}a)] and (c) effective mechanical damping $\Gamma_{\text{eff}}(\omega)$ [see Eq.~(\ref{effective}b)] versus the Fourier frequency $\omega$ in the auxiliary-cavity-unassisted ($J=0$, blue solid curves) and -assisted ($J/\omega_{m}=0.15$ and $P_{R}=50$ mW, red dashed curves) cases. The parameters are: $\Delta=\omega_{m}$, $\Delta_{a}=0$, $\omega_{m}/2\pi=10$ MHz, $\kappa_{c}/\omega_{m}=\kappa_{a}/\omega_{m}=0.1$, $\gamma_{m}/\omega_{m}=10^{-5}$, $\omega_{c}/\omega_{m}=2.817\times10^{7}$, $P_{L}=30$ mW, $m=250$ ng, $\bar{n}=10^{3}$, $L=0.5$ mm, and $\lambda=1064$ nm.}
\label{Figmodel}
\end{figure*}

In this paper, we again study how to significantly enhance the performances of refrigeration and entanglement in an auxiliary-cavity-assisted optomechanical system, revealing the robustness of the ACA entanglement against thermal noises. Inspired by optomechanical cooling and entanglement in a single-cavity setup~\cite{Wilson-Rae2007PRL,Marquardt2007PRL,Genes2008PRA,Vitali2007PRL}, we generalize
the approach for a two-cavity system, where a pumped auxiliary cavity is coupled to an optomechanical cavity. Our study differs from what is known in the double-cavity literature because we are interested not in an EIT-like mechanism~\cite{Guo2014PRA,Liu2015PRA,Gu2013PRA}, but in a pumped-auxiliary-device engineering. Using both analytical and numerical calculations based on the ACA mechanism, \emph{more than a ten-fold improvement} can be achieved for the net cooling rate, and the amplification factor almost linearly depends on the pump power of the auxiliary cavity. Physically, the ACA mechanism can significantly amplify the effective optomechanical coupling strength and considerably speed up the refrigeration process. Additionally, we show that by appropriately designing the ACA mechanism, an effective susceptibility of the mechanical resonator can be tuned largely. Moreover, by assuming experimentally reachable parameters, we find that cooling-cavity photons, auxiliary-cavity photons, and phonons can be entangled with each other, such that the steady state of the system exhibits a \emph{genuine} tripartite entanglement.

In particular, a significant enhancement can be achieved for optomechanical cooling, quantum entanglement, and noise tolerance of quantum resources. Remarkably, the cooling performance of the mechanical resonator in the ACA case is \emph{$40\%$ higher} than in cases without the auxiliary cavity. Physically, the optomechanical cooling is mainly governed by the net cooling rate, which directly determines the extraction efficiency of the thermal excitations stored in the mechanical resonator. The \emph{multiply amplified net cooling rate} leads to a significant enhancement of the cooling performance of the resonator. We find that, for the blue-detuned driving of the auxiliary cavity, both optomechanical cooling and entanglement are significantly enhanced; while for the red-detuned case, are suppressed.
Moreover, in comparison with the auxiliary-cavity-unassisted case, our ACA optomechanical entanglement has a \emph{stronger} ability against thermal noises. We also reveal that due to the joint effect of optomechanical and tunneling couplings, the \emph{indirectly coupled} cavity photons and phonons can be entangled strongly, and the robustness of quantum entanglement against thermal noises is even up to \emph{three times} that of the directly coupled case. These results provide the possibility to enhance or steer the optomechanical refrigeration and entanglement, manipulate macroscopic mechanical coherence, generate nonclassical mechanical states, as well as enhance and protect fragile quantum resources against thermal noise.

The rest of this paper is organized as follows. In Sec.~\ref{sec2}, we present the ACA optomechanical model and its Hamiltonians. In Sec.~\ref{sec3}, we derive the Langevin equations, obtain the analytical and numerical results of steady-state average phonon numbers, and calculate the logarithmic negativity. In Sec.~\ref{sec4}, we analyze the cooling performance. In Sec.~\ref{sec5}, we study bipartite and tripartite entanglements. Finally, we conclude in Sec.~\ref{sec6}. Two Appendixes include the detailed calculations of the steady-state mean phonon numbers and the bistability analysis.

\section{Model and Hamiltonian\label{sec2}}

We consider an ACA optomechanical system, where a pumped auxiliary cavity is coupled to a standard optomechanical cavity through a tunnelling coupling, as illustrated in Fig.~\ref{Figmodel}(a). A mechanical resonator is coupled to the cooling-cavity field via radiation-pressure coupling. A monochromatic laser with frequency $\omega_{L}$ ($\omega_{R}$) and field amplitude $\Omega_{L}$ ($\Omega_{R}$) is applied to drive the cooling (auxiliary) cavity, so that the optical and mechanical degrees of freedom can be manipulated. The Hamiltonian of the system reads ($\hbar =1$)
\begin{eqnarray}
\mathcal{H} &=&\omega_{c}a_{c}^{\dagger }a_{c}+\omega_{a}a_{a}^{\dagger }a_{a}+\frac{p_{x}^{2}}{2m}+\frac{m\omega _{m}^{2}x^{2}}{2}-g a_{c}^{\dagger }a_{c}x  \notag
\\&&+J(a_{c}^{\dagger }a_{a}+a_{a}^{\dagger }a_{c})+\Omega _{L}(a_{c}^{\dagger}e^{-i\omega _{L}t}+a_{c}e^{i\omega _{L}t})  \notag \\
&&+\Omega _{R}(a_{a}^{\dagger }e^{-i\omega _{R}t}+a_{a}e^{i\omega _{R}t}),\label{eq1iniH}
\end{eqnarray}
where $a_{c}$ and $a_{a}$ ($a_{c}^{\dagger }$ and $a_{a}^{\dagger }$) are the annihilation (creation) operators of the cooling-cavity and auxiliary-cavity field modes with resonance frequencies $\omega_{c}$ and $\omega_{a}$, respectively. The mechanical resonator is described by the momentum $p_{x}$ and position $x$ operators with mass $m$ and resonance frequency $\omega_{m}$. The $g$ term in Eq.~(\ref{eq1iniH}) describes the optomechanical coupling between the mechanical resonator and the cavity field, where $g=\omega_{c}/L$ is the strength of a single-photon radiation-pressure force, with $L$ being the rest length of the optical cavity.
The tunnelling coupling (with strength $J$) between the two cavity-field modes is described by the $J$ term. The last two terms in Eq.~(\ref{eq1iniH}) describe, respectively, the laser driving for the cooling and auxiliary cavities. Their amplitudes are $\Omega_{L}=\sqrt{2P_{L}\kappa_{c}/\omega_{L}}$ and $\Omega_{R}=\sqrt{2P_{R}\kappa_{a}/\omega_{R}}$, with $P_{L}$ ($P_{R}$) and $\kappa_{c}$ ($\kappa_{a}$) being the driving power and the cavity-field decay rate for the cooling (auxiliary) cavity, respectively. Note that the photon-tunneling interaction between the two cavity-field modes can be realized by optical backscattering~\cite{Jiao2020PRL,Chen2021PRL}. This backscattering of the photons is induced by the surface roughness and material defects in practical devices. Therefore, in realistic systems, the value of the photon-tunneling coupling used in our simulations should be of the same order of the decay rates of the cavity-field modes~\cite{Jiao2020PRL,Chen2021PRL}.

For convenience, we introduce the dimensionless coordinate and momentum operators $q=\sqrt{m\omega_{m}}x$ and $p=p_{x}/\sqrt{m\omega_{m}}$ ($[q,p]=i$). In a rotating frame defined by $\exp(-i\omega_{L}ta_{c}^{\dagger}a_{c}-i\omega_{R}ta_{a}^{\dagger}a_{a})$ with $\omega_{L}=\omega_{R}$, Hamiltonian~(\ref{eq1iniH}) becomes
\begin{eqnarray}
\mathcal{H}_{I} &=&\Delta _{c}a_{c}^{\dagger }a_{c}+\Delta _{a}a_{a}^{\dagger }a_{a}+\frac{\omega _{m}}{2}(q^{2}+p^{2})-g_{0}a_{c}^{\dagger }a_{c}q  \notag \\
&&+J(a_{c}^{\dagger }a_{a}+a_{a}^{\dagger }a_{c})+\Omega _{L}(a_{c}^{\dagger }+a_{c})+\Omega_{R}(a_{a}^{\dagger }+a_{a}),\label{Hamlt2dimless}
\end{eqnarray}
where $\Delta_{c}=\omega_{c}-\omega_{L}$ ($\Delta_{a}=\omega_{a}-\omega_{R}$) and $g_{0}=g/\sqrt{m\omega _{m}}$ are, respectively, the driving detuning of the cooling (auxiliary) cavity field and the strength of the optomechanical coupling expressed in terms of the dimensionless momentum and coordinate operators.

\section{Langevin equations and steady-state mean phonon numbers \label{sec3}}

In this section, we derive the quantum Langevin equations of the system and obtain the steady-state average phonon numbers in the mechanical resonator.

\subsection{Langevin equations\label{sec3A}}

To include the damping and noise effects in this system, we consider the case where the optical mode is coupled to a vacuum bath and the mechanical mode is subjected to the quantum Brownian force. In this case, the evolution of the system can be described by the quantum Langevin equations
\begin{subequations}
\label{Langevineqorig}
\begin{align}
\dot{q}=&\;\omega _{m}p, \\
\dot{p}=&-\omega _{m}q-\gamma _{m}p+g_{0}a_{c}^{\dagger }a_{c}+\xi,  \\
\dot{a_{c}}=&-[\kappa +i(\Delta _{c}-g_{0}q)]a_{c}-iJa_{a}-i\Omega +\sqrt{2\kappa_{c} }a_{c,\text{in}}, \\
\dot{a}_{a}=&-(\kappa_{a}+i\Delta_{a})a_{a}-iJa_{c}-i\Omega _{R}+\sqrt{2\kappa_{a}}a_{a\text{,in}},
\end{align}
\end{subequations}
where $\gamma_{m}$ is the decay rate of the mechanical resonator. The operators $\xi$ and $a_{c,\textrm{in}}$ ($a_{a,\textrm{in}}$), respectively, denote the Brownian force acting on the mechanical resonator and the noise operator of the cooling (auxiliary) cavity. These noise operators have zero mean values and have the following correlation functions~\cite{Genes2008PRA,Landau1958NY},
\begin{subequations}
\label{correlationfun}
\begin{align}
\langle a_{c,\textrm{in}}(t) a_{c,\textrm{in}}^{\dagger}(t^{\prime})\rangle=\delta(t-t^{\prime}), \hspace{0.5 cm}
\langle a_{c,\textrm{in}}^{\dagger}(t) a_{c,\textrm{in}}(t^{\prime})\rangle =0, \\
\langle a_{a,\textrm{in}}(t) a_{a,\textrm{in}}^{\dagger}(t^{\prime})\rangle=\delta(t-t^{\prime}), \hspace{0.5 cm}
\langle a_{a,\textrm{in}}^{\dagger}(t) a_{a,\textrm{in}}(t^{\prime})\rangle =0, \\
\langle \xi(t)\xi(t^{\prime})\rangle=\frac{\gamma_{m}}{\omega_{m}}\int e^{-i\omega(t-t^{\prime})}\omega \left[\coth\left(\frac{\omega}{2k_{B}T}\right) +1\right]\frac{d\omega }{2\pi},
\end{align}
\end{subequations}
where $k_{B}$ is the Boltzmann constant and $T$ is the reservoir temperature associated with the mechanical resonator. The correlation function in Eq.~(\ref{correlationfun}c) becomes a standard white noise
input with delta correlations for sufficiently high temperatures $k_{B}T\gg\hbar\omega_{m}$. This function can be approximated by $\langle \xi(t)\xi(t^{\prime})\rangle\approx(2\bar{n}+1)\gamma_{m}\delta(t-t^{\prime})$, where the initial mean thermal excitation number of the mechanical resonator is given by $\bar{n}=1/[\mathrm{exp}(\hbar\omega_{m}/k_{B}T)-1] \approx k_{B}T/\hbar\omega_{m}$.
To cool this mechanical resonator, we consider the strong-driving regime for both cavities, so that our physical model can be simplified by a linearization procedure. Then, we write the operators in Eq.~(\ref{Langevineqorig}) as sums of the steady-state averages and the quantum fluctuations: $o=\left\langle o\right\rangle_{\textrm{ss}} +\delta o$
for operators $a_{c}$, $a_{c}^{\dagger}$, $a_{a}$, $a_{a}^{\dagger}$, $q$, and $p$. By separating the quantum fluctuations and the classical motion, the linearized quantum Langevin equations become
\begin{subequations}
\label{fluceq}
\begin{align}
\delta \dot{q}=&\;\omega _{m}\delta p, \\
\delta \dot{p}=&-\omega _{m}\delta q-\gamma _{m}\delta p+G^{\ast }\delta a_{c}+G\delta a_{c}^{\dagger }+\xi,  \\
\delta \dot{a}_{c}=&-\bar{\kappa}_{c}\delta a_{c}+iG\delta q-iJ\delta a_{a}+\sqrt{2\kappa }a_{c,\text{in}}, \\
\delta \dot{a}_{a}=&-\bar{\kappa}_{a}\delta a_{a}-iJ\delta a_{c}+\sqrt{2\kappa _{a}}a_{a,\text{in}},
\end{align}
\end{subequations}
where $\bar{\kappa}_{c}=\kappa_{c} +i\Delta$ and $\bar{\kappa}_{a}=\kappa_{a}+i\Delta_{a}$. $\Delta =\Delta _{c}-g_{0}\langle q\rangle_{\mathrm{ss}}$ is the normalized detuning of the cooling cavity, and $G =g_{0}\langle a\rangle_{\mathrm{ss}}$ is the effective optomechanical coupling. Here $\langle a_{c}\rangle_{\mathrm{ss}}=-i(\Omega +J\langle a_{a}\rangle_{\mathrm{ss}})/(\kappa_{c} +i\Delta)$ and $\langle a_{a}\rangle_{\mathrm{ss}} =-i(\Omega _{R}+J\langle
a_{c}\rangle_{\mathrm{ss}})/(\kappa _{a}+i\Delta _{a})$. Note that we have chosen the phase reference of the cavity field, such that $\langle a_{c}\rangle_{\mathrm{ss}}$ is real and positive.

\subsection{Analytical and numerical steady-state mean phonon numbers \label{sec3C}}

Now, we derive both analytical and numerical results of the steady-state mean phonon numbers in the mechanical resonator.

\subsubsection{Analytical steady-state average phonon numbers \label{sec3C1}}

The steady-state average phonon numbers of the mechanical resonator can be obtained by the relation~\cite{Genes2008PRA,Lai2018PRA}
\begin{equation}
n_{f}=\frac{1}{2}[\langle\delta q^{2}\rangle +\langle\delta p^{2}\rangle-1],\label{finalphonumber}
\end{equation}
where $\langle\delta p^{2}\rangle$ and $\langle\delta q^{2}\rangle$ are the variances of the momentum and position operators, respectively. We obtain these variances by solving Eq.~(\ref{fluceq}) in the frequency domain, and integrating the corresponding fluctuation spectra~\cite{Genes2008PRA,Landau1958NY},
\begin{subequations}
\label{specintegral}
\begin{align}
\langle\delta q^{2}\rangle=&\;\frac{1}{2\pi}\int_{-\infty}^{\infty}S_{q}(\omega)\;d\omega,\\
\langle\delta p^{2}\rangle=&\;\frac{1}{2\pi\omega^{2}_{m}}\int_{-\infty}^{\infty}\omega^{2}S_{q}(\omega)\;d\omega,
\end{align}
\end{subequations}
where the fluctuation spectra of the momentum and position operators are defined by
\begin{equation}
S_{o}(\omega)=\int_{-\infty}^{\infty}e^{-i\omega\tau}\langle \delta o(t+\tau) \delta o(t)\rangle_{\textrm{ss}}\;d\tau,\hspace{0.5 cm}(o=q,p). \label{spectrumtimedomain}
\end{equation}
In the frequency domain, the fluctuation spectra can also be expressed as
\begin{equation}
\langle\delta\tilde{o}(\omega)\delta\tilde{o}(\omega')\rangle_{\textrm{ss}}=S_{o}(\omega) \delta(\omega+\omega').\label{spectrumfdomain}
\end{equation}
According to Eqs.~(\ref{finalphonumber}) and (\ref{specintegral}), \emph{exact results} of the steady-state average thermal excitations can be obtained analytically, which is presented in detail in the Appendix.

\subsubsection{Numerical steady-state average phonon numbers \label{sec3C2}}

Now, we introduce the annihilation (creation) operator for the mechanical resonator $b=(q+ip)/\sqrt{2}$ [$b^{\dagger}=(q-ip)/\sqrt{2}$] and then study the cooling performance by numerically evaluating the final mean phonon number. After performing the linearization, the linearized quantum Langevin equations can be rewritten as the following compact form
\begin{eqnarray}
\mathbf{\dot{u}}(t)=\mathbf{Au}(t)+\mathbf{N}(t),\label{MatrixLeq}
\end{eqnarray}
where the fluctuation operator vector $\mathbf{u}(t)=(\delta a_{c},\delta b, \delta a_{a},\delta a_{c}^{\dagger},\delta b^{\dagger}, \delta a^{\dagger}_{a})^{T}$, the noise operator vector $\mathbf{N}(t)=(\sqrt{2\kappa_{c}}a_{c,\text{in}},\sqrt{2\gamma_{m}}b_{\text{in}}, \sqrt{2\kappa_{a}}a_{a,\text{in}} ,\sqrt{2\kappa_{c}}a^{\dagger}_{c,\text{in}},\sqrt{2\gamma_{m}}b^{\dagger}_{\text{in}}, \sqrt{2\kappa_{a}}a^{\dagger}_{a,\text{in}})^{T}$, and the coefficient matrix $\mathbf{A}$:
\begin{equation}
\mathbf{A}=\left(
\begin{array}{cccccc}
-\bar{\kappa} & -i\tilde{G} & -iJ & 0 & -i\tilde{G} & 0 \\
-i\tilde{G}^{\ast } & -\bar{\gamma}_{m} & 0 & -i\tilde{G} & 0 & 0 \\
-iJ & 0 & -\bar{\kappa}_{s} & 0 & 0 & 0 \\
0 & i\tilde{G}^{\ast } & 0 & -\bar{\kappa}^{*} & i\tilde{G}^{\ast } & iJ \\
i\tilde{G}^{\ast } & 0 & 0 & i\tilde{G} & -\bar{\gamma}_{m}^{*} & 0 \\
0 & 0 & 0 & iJ & 0 & -\bar{\kappa}_{s}^{*}
\end{array}%
\right),
\end{equation}
where $\tilde{G}=G/\sqrt{2}$ and $\bar{\gamma}_{m}=\gamma _{m}+i\omega _{m}$. We then obtain the formal solution of the linearized Langevin equation~(\ref{MatrixLeq}),
\begin{equation}
\mathbf{u}(t) =\mathbf{M}(t) \mathbf{u}(0)+\int_{0}^{t}\mathbf{M}(t-s) \mathbf{N}(s)ds,\label{formal}
\end{equation}
where $\mathbf{M}(t)=\exp(\mathbf{A}t)$. From Eq.~(\ref{formal}), the steady-state mean phonon number of the mechanical resonator can be calculated by solving the Lyapunov equation. In the following calculations, all the parameters satisfy the stability conditions which are derived based on the Routh-Hurwitz criterion, i.e., the real parts of all the eigenvalues of $\mathbf{A}$ are negative. Additionally, we have confirmed that for the left pump power $P_{L}<35$ mW, only a single
stable solution exists and the compound system has no bistability (see the stability analysis in Appendix B).

Mathematically, the steady-state mean phonon number can be obtained by calculating the steady-state value of the covariance matrix $\mathbf{V}$, defined by the matrix elements
\begin{equation}
\mathbf{V}_{ij}=\frac{1}{2}[\langle \mathbf{u}_{i}(\infty) \mathbf{u}_{j}(\infty ) \rangle +\langle \mathbf{u}_{j}( \infty) \mathbf{u}_{i}(\infty )\rangle], \hspace{0.5 cm}i,j=1-6.
\end{equation}
Under the stability conditions, the steady-state covariance matrix $\mathbf{V}$ fulfills the Lyapunov equation
\begin{equation}
\mathbf{A}\mathbf{V}+\mathbf{V}\mathbf{A}^{T}=-\mathbf{Q}, \label{Lyapunov}
\end{equation}
where the superscript $T$ represents transposition and
\begin{equation}
\mathbf{Q}=\frac{1}{2}(\mathbf{C}+\mathbf{C}^{T}),
\end{equation}
with $\mathbf{C}$ being the noise correlation matrix defined by the matrix elements
\begin{eqnarray}
\langle \mathbf{N}_{k}(s) \mathbf{N}_{l}(s^{\prime})\rangle =\mathbf{C}_{k,l}\delta (s-s^{\prime }).
\end{eqnarray}
For the Markovian bath considered in our work, the constant matrix $\mathbf{C}$ is expressed as
\begin{equation}
\mathbf{C}=\left(
\begin{array}{cccccc}
0 & 0 & 0 & 2\kappa & 0 & 0\\
0 & 0 & 0 & 0 & 2\gamma_{m}(\bar{n}+1) & 0\\
0 & 0 & 0 & 0 & 0 & 2\kappa_{s}\\
0 & 0 & 0 & 0 & 0 & 0\\
0 & 2\gamma _{m}\bar{n} & 0 & 0 & 0 & 0\\
0 & 0 & 0 & 0 & 0 & 0
\end{array}
\right).
\end{equation}
By calculating the covariance matrix $\mathbf{V}$, we obtain the steady-state mean phonon number
\begin{align}
n_{f}=\langle \delta b^{\dagger}\delta b\rangle=\mathbf{V}_{52}-\frac{1}{2},\label{finalexact}
\end{align}
where $\mathbf{V}_{52}$ is obtained by solving the Lyapunov equation ~(\ref{Lyapunov}).

\section{ACA optomechanical cooling\label{sec4}}

In this section, we study the ACA cooling by analyzing the effective mechanical susceptibility, the net laser-cooling rate, and the noise spectra.

\subsection{Analytical results of the effective susceptibility and net cooling rate \label{sec4A}}

We obtain the position fluctuation spectrum of the mechanical resonator as
\begin{eqnarray}
S_{q}(\omega )&=&|\chi _{\text{eff}}(\omega )|^{2}\Big[S_{\text{rp}}(\omega )+S_{\text{th}}(\omega )\Big],\label{spectra12}
\end{eqnarray}
where $\chi_{\text{eff}}(\omega)$ is the \emph{effective susceptibility} of the mechanical resonator, given by
\begin{equation}
\chi_{\text{eff}}(\omega)=\omega_{m}[\Omega_{\text{eff}}^{2}(\omega )-\omega ^{2}-i\omega \Gamma _{\text{eff}}(\omega )]^{-1},\label{susceptibility}
\end{equation}
with $\Omega_{\text{eff}}(\omega)$ and $\Gamma_{\text{eff}}(\omega )$ being, respectively, the \emph{effective resonance frequency} and \emph{damping rate} of the mechanical resonator, defined as
\begin{subequations}
\label{effective}
\begin{eqnarray}
\Omega _{\text{eff}} &=&\sqrt{\omega _{m}^{2}-2\left\vert G\right\vert
^{2}\omega _{m}(\varphi \Pi +2\Delta \kappa _{a}\omega ^{2}\Phi )/\zeta}, \\
\Gamma _{\text{eff}} &=&\gamma _{m}+\gamma _{\text{C}}.
\end{eqnarray}
\end{subequations}
Here, $\gamma _{\text{C}}$ denotes the \emph{net cooling rate} of the mechanical resonator, defined as
\begin{eqnarray}
\gamma _{\text{C}}=2\left\vert G\right\vert ^{2}\omega _{m}(2\Delta\kappa _{a}\Pi -\varphi \Phi )/\zeta,\label{coolingrate0}
\end{eqnarray}
and other parameters are
\begin{subequations}
\label{coeff01}
\begin{eqnarray}
\Pi  &=&\beta _{+}\beta _{-}+\tau _{+}\tau _{-}, \\
\Phi  &=&2[\kappa_{c} ^{2}\kappa _{a}+J^{2}(\kappa_{c} +\kappa _{a})+\kappa_{a}(\Delta ^{2}-\omega ^{2}) \notag \\
&&+\kappa_{c} (\kappa _{a}^{2}-\omega ^{2}+\Delta _{a}^{2})], \\
\zeta  &=&(\beta _{+}^{2}+\tau _{+}^{2})(\beta _{-}^{2}+\tau _{-}^{2}), \\
\varphi  &=&J^{2}\Delta _{a}-\Delta (\kappa _{a}^{2}-\omega ^{2}+\Delta_{a}^{2}),
\end{eqnarray}
\end{subequations}
with
\begin{subequations}
\label{coeff0}
\begin{align}
\beta _{\pm } =&\pm J^{2}\pm \kappa_{c} \kappa _{a}\mp (\omega \pm \Delta)(\omega \pm \Delta _{a}), \\
\tau _{\pm } =&\kappa_{c} (\omega \pm \Delta _{a})+\kappa _{a}(\omega \pm\Delta ).
\end{align}
\end{subequations}
In Eq.~(\ref{spectra12}), the thermal noise spectrum $S_{\text{th}}(\omega )$ is given by,
\begin{eqnarray}
S_{\text{th}}(\omega) &=&\frac{\gamma_{m}\omega }{\omega _{m}}\coth
\left( \frac{\hbar \omega }{2\kappa _{B}T}\right),\label{Spectra1}
\end{eqnarray}
and the radiation-pressure noise spectrum $S_{\text{rp}}(\omega )$ is so complicated that we don't show it here.

\subsection{Amplified net cooling rate \label{sec4AB}}

\begin{figure}[tbp]
\center
\includegraphics[ width=0.48 \textwidth]{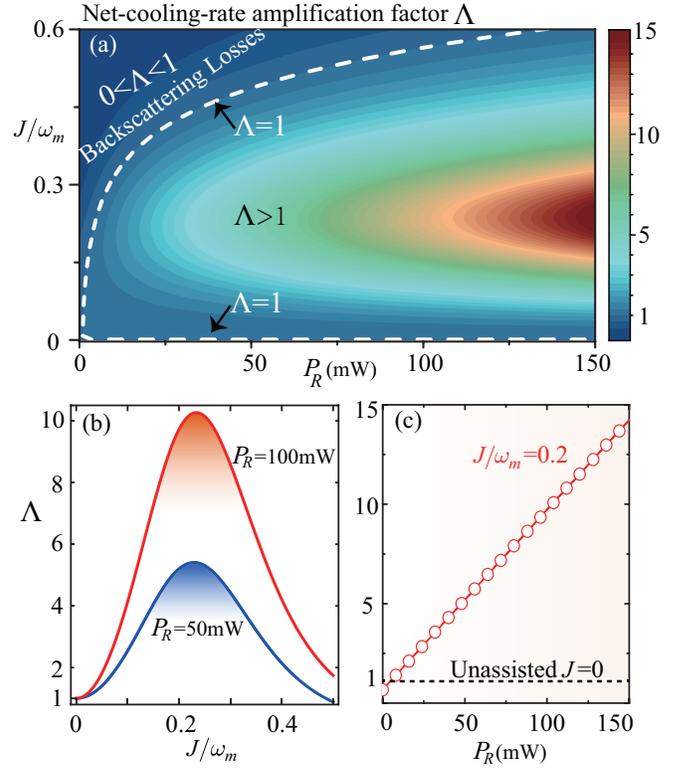}
\caption{ (a) Net-cooling-rate amplification factor $\Lambda$ [see Eq.~(\ref{amplify})] versus the tunneling coupling $J$ and the drive laser power $P_{R}$ of the auxiliary cavity. The white dashed curve denotes $\Lambda=1$. (b) $\Lambda$ versus $J$ when $P_{R}=50$ mW (blue solid curve) and $P_{R}=100$ mW (red solid curve). (c) $\Lambda$ versus $P_{R}$ when $J=0$ (black horizontal dashed line) and  $J/\omega_{m}=0.2$ (red symbols). Here the black horizontal dashed line denotes the auxiliary-cavity-unassisted case, i.e, $J=0$. Other parameters are the same as those used in Fig.~\ref{Figmodel}.}
\label{amplification}
\end{figure}

\begin{figure}[tbp]
\center
\includegraphics[width=0.47 \textwidth]{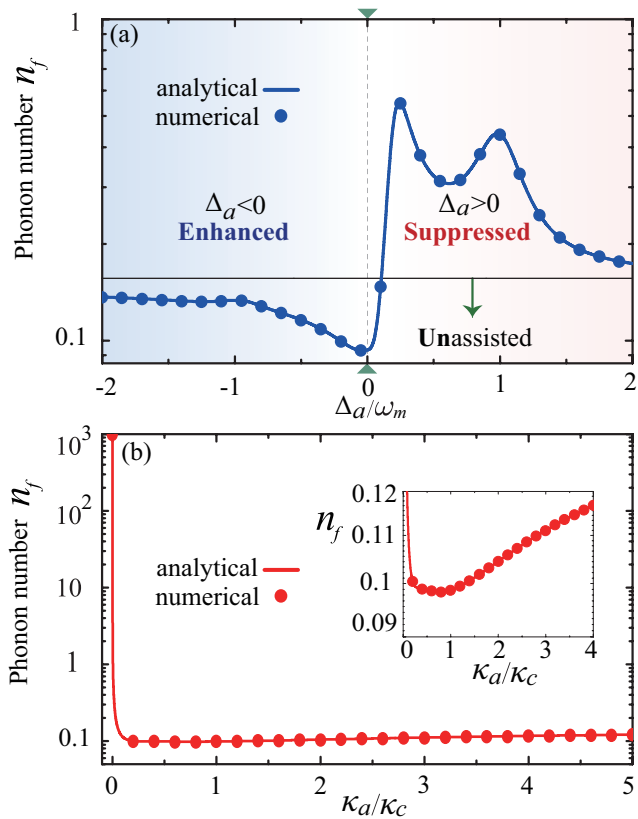}
\caption{(a) Steady-state average phonon number $n_{f}$ versus the effective driving detuning $\Delta_{a}$ when $\kappa_{a}/\kappa_{c}=1$. Curves show our analytical predictions and symbols are the numerical results. Here the black horizontal solid line denotes the auxiliary-cavity-unassisted case, i.e, $J=0$. (b) Steady-state average phonon number $n_{f}$ as a function of $\kappa_{a}$ when $\Delta_{a}=0$. Other parameters are the same as those used in Fig.~\ref{Figmodel}.}
\label{Detasks}
\end{figure}

In the preceding subsection, the effective mechanical resonance frequency $\Omega _{\text{eff}}$, mechanical damping rate $\Gamma_{\text{eff}}$ [see Eq.~(\ref{effective})], and the expression of the steady-state average thermal excitation [see Eq.~(\ref{exactcoolresult})] have been analyzed analytically. Now, we study how the ACA mechanism improves the cooling performance by analyzing the effective decay rate $\Gamma_{\text{eff}}$ and mechanical resonance frequency $\Omega_{\text{eff}}$.

\begin{figure*}[tbp]
\center
\includegraphics[width=1 \textwidth]{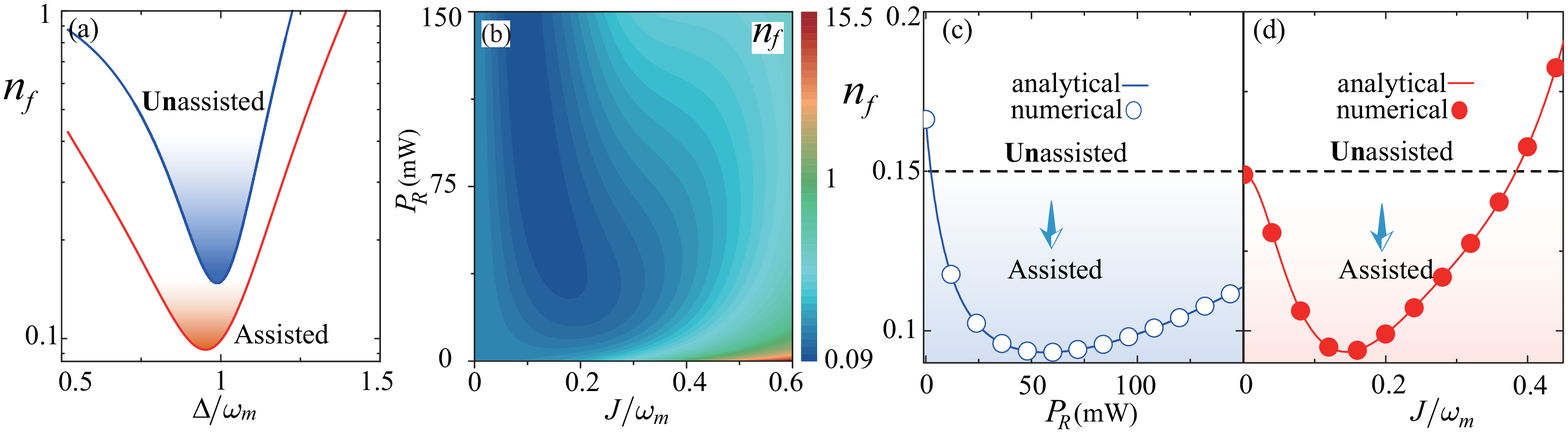}
\caption{(a) Steady-state average phonon number $n_{f}$ versus the effective driving detuning $\Delta$ in the auxiliary-cavity-unassisted ($J=0$, blue solid curve) and auxiliary-cavity-assisted ($J/\omega_{m}=0.15$ and $P_{R}=50$ mW, red solid curve) cases. (b) $n_{f}$ as a function of the tunneling coupling strength $J$ and the laser power $P_{R}$ of the auxiliary cavity when $\Delta=\omega_{m}$. Based on the analytical [see Eq.~(\ref{exactcoolresult}), solid curves] and numerical [see Eq.~(\ref{finalexact}), symbols] results, we plot (c) $n_{f}$ versus $P_{R}$ when $J/\omega_{m}=0.15$; and (d) $n_{f}$ versus $J$ when $P_{R}=50$ mW. Here the black horizontal dashed lines denote the auxiliary-cavity-unassisted case, i.e, $J=0$. Other parameters used are the same as those in Fig.~\ref{Figmodel}.}
\label{phonon}
\end{figure*}

When the system works in the auxiliary-cavity-unassisted ($J=0$, see solid curves) and -assisted ($J/\omega_{m}=0.2$ and $P_{R}=30$ mW, see dashed curves) cases, we plot the effective mechanical resonance frequency $\Omega_{\text{eff}}$ and decay rate $\Gamma_{\text{eff}}$ as a function of the frequency $\omega$, as shown in Figs.~\ref{Figmodel}(b) and ~\ref{Figmodel}(c). We find that the modification of the mechanical frequency, mainly determined by the optomechanical coupling shown in Eq.~(\ref{effective}a), is the so-called ``optical spring effect", which may lead to significant frequency shifts in the case of low-frequency mechanical resonators. However, for our higher-resonance frequency ($\omega_{m}/2\pi=10$ MHz), the optical spring term in Eq.~(\ref{effective}a) does not significantly alter the mechanical frequency, i.e., $\Omega_{\text{eff}}\approx\omega_{m}$ when $\omega/\omega_{m}=\pm1$ [see Fig.~\ref{Figmodel}(b)]. Moreover, we show in Fig.~\ref{Figmodel}(c) that by using the ACA mechanism, $\Gamma_{\text{eff}}$ is significantly increased at $\omega=\pm\omega_{m}$. For example, when we switch the unassisted to assisted cases, the effective mechanical decay rate $\Gamma_{\text{eff}}$ at $\omega=\pm\omega_{m}$ can be increased from $\approx10^{4}\gamma_{m}$ to $\approx4.5\times10^{4}\gamma_{m}$. This giant enhancement of the effective mechanical damping $\Gamma_{\text{eff}}$ plays an important role in improving the cooling performance of the mechanical resonator.

To further understand the underlying physics of the ACA cooling, we consider the red-sideband resonance case, i.e., $\Delta=\omega_{m}$ and $\omega=\omega_{m}$, and then define a net-cooling-rate amplification factor
\begin{eqnarray}
\Lambda=\frac{\gamma_{C,\mathrm{assisted}}}{\gamma_{C,\mathrm{unassisted}}}.\label{amplify}
\end{eqnarray}
In Fig.~\ref{amplification}(a), we plot the net-cooling-rate amplification factor $\Lambda$ with respect to the tunneling coupling $J$ and the pump power $P_{R}$ of the auxiliary cavity. It shows that the ACA method can significantly amplify the net cooling rate of the mechanical resonator. For example, in the unassisted case (i.e., when $J=0$), there is no cooling-rate amplification (i.e., $\Lambda=1$), while in the assisted case, the amplification of the net cooling rate emerges and even the amplification factor can increase up to $\Lambda=15$. In particular, when $P_{R}\rightarrow0$ and $J/\omega_{m}\rightarrow0.6$, we obtain $0<\Lambda<1$ [see the upper left corner in Fig.~\ref{amplification}(a)], which is due to the optical backscattering losses in practical devices. Physically, various imperfections of devices, such as material
defects and surface roughness, can induce backscattering of photons, as described by the tunneling coupling $J$~\cite{Jiao2020PRL,Chen2021PRL}. In a recent experiment~\cite{Kim2019OP}, a dynamical suppression of backscattering was already observed by breaking the time-reversal symmetry with Brillouin devices. The dependence of the net-cooling-rate amplification factor $\Lambda$ on the tunneling coupling $J$ between the two cavities is shown in Fig.~\ref{amplification}(b). We find that in the region $0<J/\omega_{m}<0.25$ ($0.25<J/\omega_{m}<0.5$), $\Lambda$ increases (decreases) with increasing $J$, and the optimal amplification factor emerges at $J/\omega_{m}=0.25$. Particularly, in comparison with the typical optomechanical systems, a proportional amplification of the net cooling rate can be observed with the pump power $P_{R}$ when $J\approx0.2\omega_{m}$ [see Figs.~\ref{amplification}(c)]. Physically, the effective optomechanical coupling strength can be amplified and the refrigeration process can be accelerated, by utilizing the ACA mechanism. This study provides a new strategy to improve the net cooling rate of the mechanical resonator by just using a pumped auxiliary device.

\subsection{ACA optomechanical cooling\label{sec4B}}

The foremost task of studying cooling properties in such an ACA optomechanical system is to find the optimal driving detuning $\Delta_{a}$ and decay rate $\kappa_{a}$ of the pumped auxiliary cavity. In Fig.~\ref{Detasks}(a) we show the steady-state average phonon numbers $n_{f}$ of the mechanical resonator versus driving detuning $\Delta_{a}$ of the pumped auxiliary cavity. We find a significant enhancement for the cooling performance for blue-detuned driving, $\Delta_{a}<0$, and that the optimal cooling is located at $\Delta_{a}=0$. In contrast, the red-detuned driving, $\Delta_{a}>0$, leads to the suppression of the cooling efficiency. Here, the black horizontal solid line denotes the auxiliary-cavity-unassisted case, i.e, $J=0$. Additionally, we plot the steady-state mean phonon numbers $n_{f}$ as a function of the decay rate $\kappa_{a}$ of the auxiliary cavity, as shown in Fig.~\ref{Detasks}(b). We can see that the optimal cooling efficiency of the mechanical resonator emerges in $0.5<\kappa_{a}/\kappa_{c}<1$. Note that our numerical (marked by symbols) and analytical (solid curves) results exhibit an excellent agreement, as shown in Fig.~\ref{Detasks}. These results indicate a large improvement of the cooling performance, which is realized by an appropriate design of the auxiliary cavity.

In Fig.~\ref{phonon}(a) the final average phonon numbers $n_{f}$ are plotted as a function of the effective driving detuning $\Delta$ of the cooling cavity when the system works in both the auxiliary-cavity-unassisted (see the blue curve) and -assisted (see the red curve) regimes. We can see that when the system is in the assisted case, the cooling performance is much better than that in the unassisted case [see Fig.~\ref{phonon}(a)]. This is because the use of the pumped auxiliary cavity can significantly amplify the net-cooling rate of the mechanical resonator and, then, considerably improve its refrigeration performance. Note that for the unassisted case, the mechanical resonator is cooled in the same manner as in a typical optomechanical sideband-cooling scheme~\cite{Wilson-Rae2007PRL,Marquardt2007PRL,Genes2008PRA}. The optimal driving detuning is located at $\Delta\approx\omega_{m}$, which indicates the maximum energy extraction efficiency between the cooling-cavity-field mode and the mechanical resonator.

In realistic simulations, we find a small deviation of the exact value of $\omega_{m}$. This is caused by the counter-rotating-wave term in the linearized coupling between the cooling-cavity field and the mechanical resonator. The underlying physics is that the generation of an anti-Stokes photon leads to the cooling of the mechanical resonator by taking away a phonon from this resonator. For the optimal
cooling $\Delta\approx\omega_{m}$, the frequency $\omega_{m}$ of the phonon exactly matches the driving detuning $\Delta$, and hence $\Delta\approx\omega_{m}$ corresponds to the optimal cooling.

To further elucidate this cooling improvement, we plot the final average phonon numbers $n_{f}$ as functions of the tunneling coupling $J$ and the pump power $P_{R}$ of the auxiliary cavity, as shown in Fig.~\ref{phonon}(b). By using the ACA mechanism, the mechanical resonator can be cooled efficiently ($n_{f}\ll1$), and the lowest final average occupancies are $0.09$, which is much smaller than that of the auxiliary-cavity-unassisted case. We can see from Figs.~\ref{phonon}(c) and~\ref{phonon}(d) that the cooling performance is fully unchanged (see the black dashed lines, $n_{f}=0.15$) in the unassisted case, but in stark contrast is improved strongly (see the solid curves and symbols, $n_{f}=0.09$) in the assisted case. Note that the numerical (symbols) and analytical (solid curves) results show an excellent agreement, as seen in Figs.~\ref{phonon}(c) and~\ref{phonon}(d).

\begin{figure}[tbp]
\center
\includegraphics[width=0.47 \textwidth]{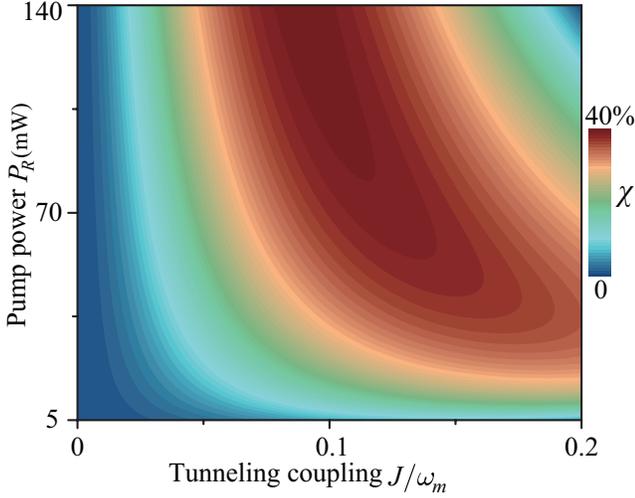}
\caption{Cooling-performance improvement rate $\chi$ of the mechanical resonator [see Eq.~(\ref{improved})] versus the tunneling coupling strength $J$ and the drive power $P_{R}$ of the auxiliary cavity. Other parameters are the same as those used in Fig.~\ref{Figmodel}.}
\label{improverate}
\end{figure}

To study how large the significant improvement of the cooling performance can be reached, we here introduce a cooling-performance improvement rate $\chi$, defined as
\begin{eqnarray}
\chi=\frac{n_{f,\mathrm{assisted}}-n_{f,\mathrm{unassisted}}}{n_{f,\mathrm{unassisted}}}.\label{improved}
\end{eqnarray}
Based on Eq.~(\ref{improved}), we investigated the dependence of the cooling-performance improvement rate $\chi$ on the parameters $J$ and $P_{R}$, as shown in Fig.~\ref{improverate}. We can see for the ACA system, that the rate $\chi$ can reach $40\%$ compared with the unassisted case, and that the optimal cooling performance is for: $0.07<J/\omega_{m}<0.2$ and $35<P_{R}<140$ mW. Physically, the optomechanical cooling is mainly governed by the net cooling rate, which directly determines the extraction efficiency of thermal excitations stored in the mechanical resonator and, therefore, the greatly amplified net cooling rate leads to a significant improvement of the cooling performance.

In the above simulations, we have found that the optimal cooling performance is observed at $\Delta\approx\omega_{m}$ [see Fig.~\ref{Detasks}(a)], corresponding the maximum phonon extraction efficiency. Thus, based on our analytical expression in Eq.~(\ref{exactcoolresult}) of the final mean phonon number, the analytical result of the minimum occupation number can be achieved by setting $\Delta=\omega_{m}$ and $\gamma_{m}=0$. However, the analytical expression of the minimum occupation number is so complicated that we do not show it here. Below, we study this by numerical simulations.

\begin{figure}[tbp]
\center
\includegraphics[width=0.47 \textwidth]{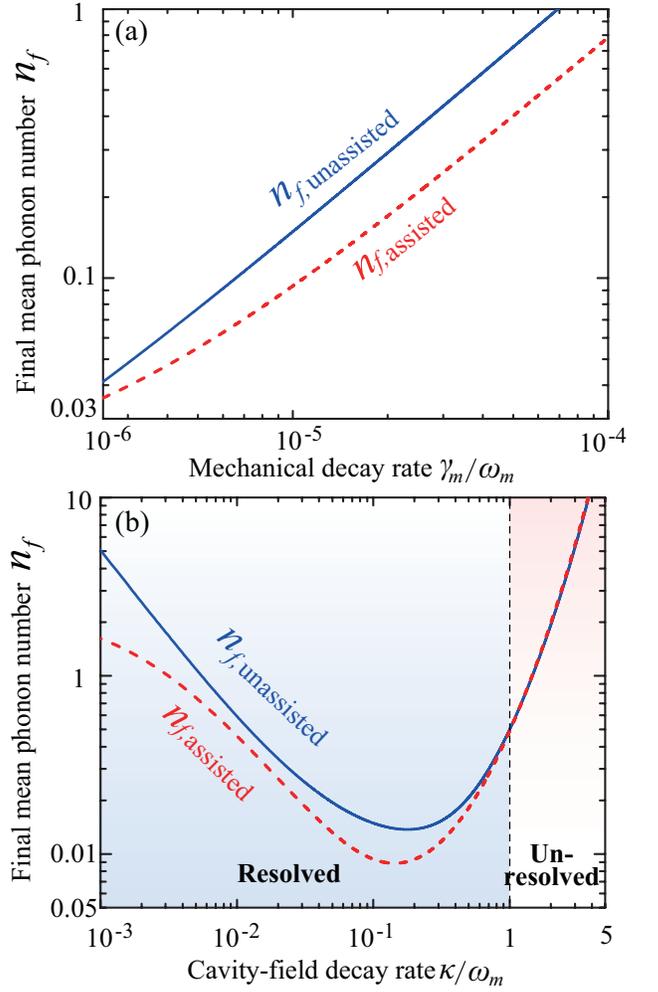}
\caption{Steady-state average phonon number $n_{f}$ versus (a) the mechanical decay rate $\gamma_{m}$ when $\kappa_{c}/\omega_{m}=0.1$, and (b) the cavity-field decay $\kappa_{c}$ when $\gamma_{m}/\omega_{m}=10^{-5}$, in the auxiliary-cavity-unassisted ($n_{f,\mathrm{unassisted}}$, blue solid curves) and auxiliary-cavity-assisted ($n_{f,\mathrm{assisted}}$, red dashed curves) cases. Here we take the optimal effective driving detuning $\Delta=\omega_{m}$. Other parameters are the same as those used in Fig.~\ref{Figmodel}.}
\label{decayrate}
\end{figure}

Under the optimal effective driving detuning, $\Delta=\omega_{m}$, we plot the steady-state average phonon number $n_{f}$ as a function of the mechanical decay rate $\gamma_{m}$ in the auxiliary-cavity-unassisted (blue solid curve) and auxiliary-cavity-assisted (red dashed curve) cases, as shown in Fig.~\ref{decayrate}(a). We find that in both unassisted and assisted cases, the redundant single-phonon probability could be further suppressed by choosing the mechanical resonator with a smaller decay rate. Physically, the thermal phonon extraction rate (between the mechanical resonator and its heat bath) is faster for a larger value of the mechanical decay rate, and, then, the thermal excitations in the heat bath increase the phonon numbers in the mechanical resonator. In particular, we observe that the cooling efficiency of the mechanical resonator in the assisted case is higher than that of the unassisted case (i.e., $n_{f,\mathrm{assisted}}<n_{f,\mathrm{unassisted}}$), and that the minimum occupation number of the resonator in the assisted case is smaller than that in the unassisted case when $\gamma_{m}\rightarrow0$. This is because our ACA mechanism can significantly amplify the effective optomechanical coupling strength and considerably improve the refrigeration performance.

In Fig.~\ref{decayrate}(b), the final steady-state mean phonon number $n_{f}$ is plotted as a function of the cavity-field decay rate $\kappa$, when the system operates in both unassisted and assisted cases. To clearly study the influence of the sideband-resolution condition on the cooling performance, we also choose the mechanical frequency $\omega_{m}$ as the frequency scale. We can see that in both unassisted and assisted cases, the phonon sidebands can be well resolved from the cavity-emission spectrum when $\kappa_{c}/\omega_{m}\ll1$ [see the left area of the dashed back line in Fig.~\ref{decayrate}(b)], which is called the resolved-sideband limit. In the unresolved-sideband regime  $\kappa_{c}/\omega_{m}>1$, the cooling performance of the mechanical resonator becomes much worse for a larger cavity-field decay rate $\kappa_{c}$ [see the right area of the dashed back line in Fig.~\ref{decayrate}(b)]. This is due to the decrease of the net-cooling rate $\gamma_{\mathrm{C}}$. In particular, the optimal cooling performance is observed for $\kappa_{c}/\omega_{m}\approx0.1-0.3$, and that the cooling performance of the assisted case is better than that of the unassisted case (i.e., $n_{f,\mathrm{assisted}}<n_{f,\mathrm{unassisted}}$). When $\kappa_{c}/\omega_{m}<0.1$, the cooling performance becomes worse with the decrease of $\kappa_{c}$, because the net-cooling rate $\gamma_{\mathrm{C}}\rightarrow0$ when $\kappa_{c}/\omega_{m}\rightarrow0$~\cite{Li2008PRB}. Physically, the thermal excitations stored in the mechanical resonator are mainly first transferred to the cavity and then leak from the cavity through the bath coupled to the cavity. When $\kappa_{c}/\omega_{m}\rightarrow0$, the thermal energy leakage from the cavity is too weak and one could not obtain a strong cooling. These findings provide a method to develop the cooling performance of the mechanical resonator by appropriately designing a auxiliary device.

\section{ACA quantum entanglement and its noise-tolerant ability\label{sec5}}

Now we study the effect of the ACA cooling mechanism on quantum entanglement and its robustness against thermal noises by calculating the logarithmic negativity.

\subsection{Logarithmic negativity and minimum residual contangle\label{sec5A}}

Let us define the quadrature fluctuations with $\delta X_{o}=(\delta o^{\dagger}+\delta o)/\sqrt{2}$ and $\delta Y_{o}=i(\delta o^{\dagger}-\delta o)/\sqrt{2}$ for $o\in$\{ $a_{c}$, $a_{c}^{\dagger}$, $a_{a}$, $a_{a}^{\dagger}$, $q$, $p$\}, and the corresponding Hermitian input noise operators with $\delta X^{\mathrm{in}}_{o}=(o_{\mathrm{in}}^{\dagger}+o_{\mathrm{in}})/\sqrt{2}$ and $\delta Y^{\mathrm{in}}_{o}=i(o_{\mathrm{in}}^{\dagger}-o_{\mathrm{in}})/\sqrt{2}$. Then, the linearized equations of fluctuations can be written as
\begin{eqnarray}
\mathbf{\dot{\tilde{u}}}(t)=\mathbf{\tilde{A}\tilde{u}}(t)+\mathbf{\tilde{N}}(t),\label{MatrixLeq2}
\end{eqnarray}
where $\mathbf{\tilde{u}}(t)=[\delta X_{a_{c}}, \delta Y_{a_{c}}, \delta X_{a_{a}}, \delta Y_{a_{a}},
\delta q, \delta p]^{T}$ is the vector of fluctuation operators, $\mathbf{\tilde{N}}(t)=\left( \sqrt{2\kappa }X_{a_{c}}^{\text{in}},\sqrt{2\kappa }%
Y_{a_{c}}^{\text{in}},\sqrt{2\kappa _{a}}X_{a_{a}}^{\text{in}},\sqrt{2\kappa
_{a}}Y_{a_{a}}^{\text{in}},0,\xi \right)^{T}$ is the vector of input noises, and the coefficient matrix $\mathbf{A}$ is given by
\begin{equation}
\mathbf{\tilde{A}}=\left(
\begin{array}{cccccc}
-\kappa _{c} & \Delta  & 0 & J & 0 & 0 \\
-\Delta  & -\kappa _{c} & -J & 0 & \sqrt{2}G & 0 \\
0 & J & -\kappa _{a} & \Delta _{a} & 0 & 0 \\
-J & 0 & -\Delta _{a} & -\kappa _{a} & 0 & 0 \\
0 & 0 & 0 & 0 & 0 & \omega _{m} \\
\sqrt{2}G & 0 & 0 & 0 & -\omega _{m} & -\gamma _{m}%
\end{array}%
\right).
\end{equation}%
The formal solution of Eq.~(\ref{MatrixLeq2}) is $\mathbf{\tilde{u}}(t) =\mathbf{\tilde{M}}(t) \mathbf{\tilde{u}}(0)+\int_{0}^{t}\mathbf{\tilde{M}}(t-s)\mathbf{\tilde{N}}(s)ds$, where $\mathbf{\tilde{M}}(t)=\exp(\mathbf{\tilde{A}}t)$. Now we can calculate the steady-state value of the covariance matrix $\mathbf{\tilde{V}}$, which is defined by the matrix elements $\mathbf{\tilde{V}}_{kl}=\frac{1}{2}[\langle \mathbf{\tilde{u}}_{k}(\infty) \mathbf{\tilde{u}}_{l}(\infty ) \rangle +\langle \mathbf{\tilde{u}}_{l}( \infty) \mathbf{\tilde{u}}_{k}(\infty )\rangle]$, for $k,l=1$-$6$. Under the stability condition, the covariance matrix $\mathbf{\tilde{V}}$ fulfills the Lyapunov equation
$\mathbf{\tilde{A}}\mathbf{\tilde{V}}+\mathbf{\tilde{V}}\mathbf{\tilde{A}}^{T}=-\mathbf{\tilde{Q}}$,
where $\mathbf{\tilde{Q}}=\mathrm{diag} \{\kappa_{c},\kappa_{c},\kappa_{a},\kappa_{a},0,\gamma_{m}(2\bar{n}+1)\}$. To study the bipartite entanglement of the system, we adopt quantitative measures of the logarithmic negativity $E_{\mathcal{N}}$, defined as~\cite{Vidal2002PRA,Plenio2005PRL,Adesso2004PRA}
\begin{align}
E_{\mathcal{N}}=\mathrm{max}[0,-\mathrm{ln}(2\zeta^{-})],
\end{align}
where $\zeta^{-}\equiv 2^{-1/2}\{\Sigma(\mathbf{\tilde{V}}^{'})-[\Sigma(\mathbf{\tilde{V}}^{'})^{2}-4 \mathrm{det} \mathbf{\tilde{V}}^{'}]^{1/2}\}^{1/2}$, with $\Sigma(\mathbf{\tilde{V}}^{'})\equiv\mathrm{det}\mathcal{A}+\mathrm{det}\mathcal{B}-2\mathrm{det}\mathcal{C}$. Here the matrix $\mathbf{\tilde{V}}^{'}$ is written as
\begin{align}
\mathbf{\tilde{V}}^{'}=&\left(
\begin{array}{cc}
\mathcal{A} & \mathcal{C} \\
\mathcal{C}^{T} & \mathcal{B}%
\end{array}%
\right),
\end{align}
where $\mathcal{A}$, $\mathcal{B}$, and $\mathcal{C}$ are $2\times2$ subblock matrices of $\mathbf{\tilde{V}}^{'}$.

\begin{figure*}[tbp]
\center
\includegraphics[width=1 \textwidth]{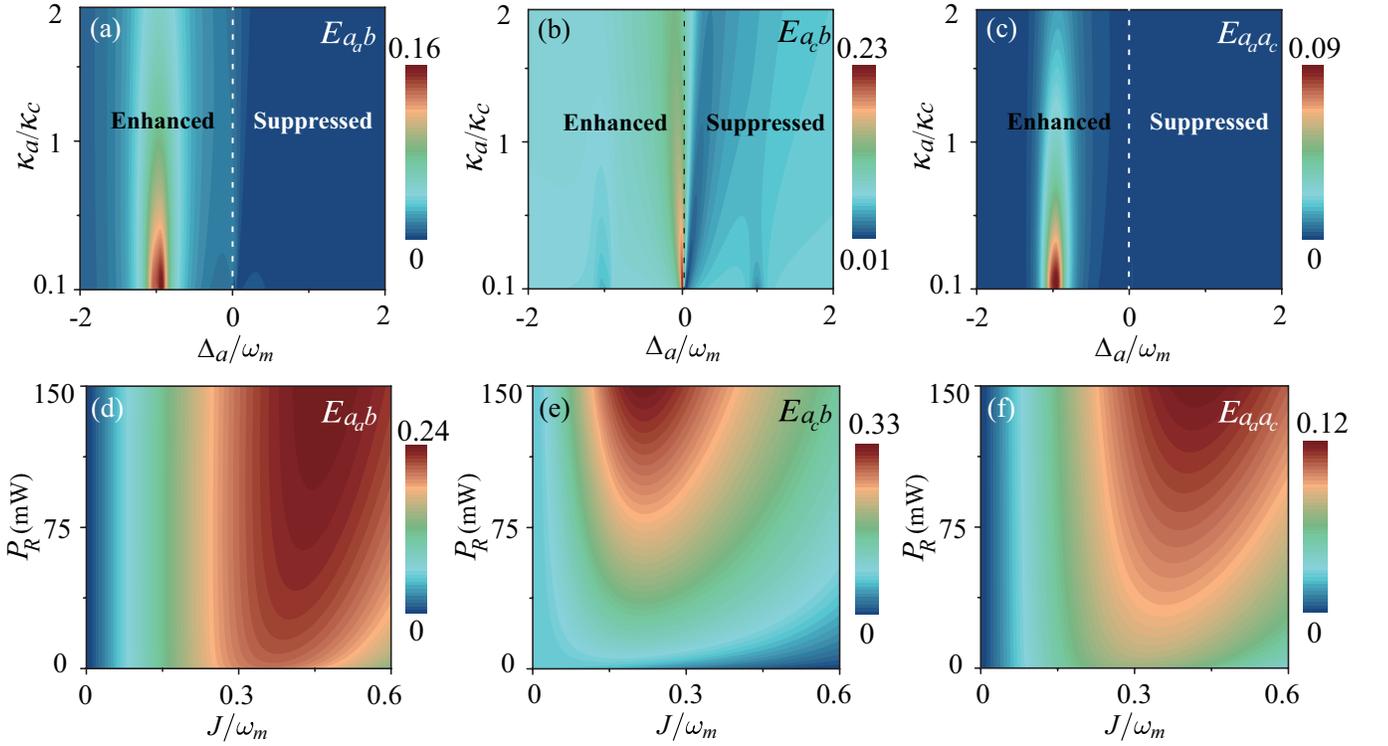}
\caption{Logarithmic negativities (a) $E_{a_{a}b}$, (b) $E_{a_{c}b}$, and (c) $E_{a_{a}a_{c}}$ versus the driving detuning $\Delta_{a}$ and the decay rate $\kappa_{a}$ of the auxiliary cavity in the ACA case. (d) $E_{a_{a}b}$, (e) $E_{a_{c}b}$, and (f) $E_{a_{a}a_{c}}$ versus the tunneling coupling $J$ and the drive power $P_{R}$ of the auxiliary cavity when $\kappa_{a}=0.5\kappa_{c}$ and $\Delta_{a}/\omega_{m}=-1$ (for $E_{a_{a}b}$ and $E_{a_{a}a_{c}}$) and $=0$ (for $E_{a_{c}b}$). Other parameters are the same as those used in Fig.~\ref{Figmodel}.}
\label{Entanglement}
\end{figure*}

For studying the tripartite entanglement of the system, we first apply a quantitative measure of the residual contangle $\bar{E}^{r|s|t}_{\tau}$~\cite{Adesso2007JPA,Li2018PRL,Coffman2000PRA}, which is given by
\begin{align}
\bar{E}^{r|s|t}_{\tau}\equiv&\;E^{r|(st)}_{\tau}- E^{r|s}_{\tau}-E^{r|t}_{\tau},~~~(r,s,t=d_{1},d_{2},c),
\end{align}
where
$E^{u|v}_{\tau}$ denotes the contangle of subsystems $u$ ($u$ contains only one mode) and $v$ ($v$ contains one or two modes). $E^{u|v}_{\tau}$ is a proper entanglement monotone and it can be defined as the \emph{squared} logarithmic negativity~\cite{Adesso2007JPA,Li2018PRL,Coffman2000PRA}.
The residual contangle satisfies the monogamy property of quantum entanglement, $E^{r|(st)}_{\tau}\geq0$, i.e.,
\begin{eqnarray}
E^{r|(st)}_{\tau}\geq E^{r|s}_{\tau}+E^{r|t}_{\tau}.
\end{eqnarray}
This inequality is analogous to the popular Coffman-Kundu-Wootters monogamy inequality, which holds for three qubits~\cite{Coffman2000PRA}.

A bona fide quantification of continuous-variable tripartite entanglement is provided by the \emph{minimum} residual contangle~\cite{Adesso2007JPA,Li2018PRL,Coffman2000PRA}
\begin{eqnarray}
E^{r|s|t}_{\tau}\equiv& \min \limits_{(r,s,t)}[E^{r|(st)}_{\tau}-E^{r|s}_{\tau}-E^{r|t}_{\tau}],\label{minimum}
\end{eqnarray}
where $(r,s,t)\equiv(d_{1},d_{2},c)$ denotes all the permutations of the three mode indexes~\cite{Adesso2007JPA}. The nonzero minimum residual contangle $E^{r|s|t}_{\tau}>0$ means that the \emph{genuine} tripartite entanglement is generated.

\subsection{ACA bipartite entanglements\label{sec5B}}

To study quantum entanglement properties of this system, the foremost task is to find the optimal detuning $\Delta_{a}$ and decay rate $\kappa_{a}$ of the pumped auxiliary cavity.
In Figs.~\ref{Entanglement}(a),~\ref{Entanglement}(b), and~\ref{Entanglement}(c), we present a quantum entanglement measure, i.e., the logarithmic negativity, versus the driving detuning $\Delta_{a}$ and decay rate $\kappa_{a}$: $E_{a_{a}b}$, $E_{a_{c}b}$, and $E_{a_{a}a_{c}}$ are the auxiliary-cavity-phonon, cooling-cavity-phonon, and photon-photon entanglements, respectively. Note that all the parameters satisfy the stability conditions, which are derived from the Routh-Hurwitz criterion, i.e., the real parts of all the eigenvalues of $\mathbf{\tilde{A}}$ are negative. We can see from Figs.~\ref{Entanglement}(a) and~\ref{Entanglement}(c) that, for the red-detuned driving of the auxiliary cavity, i.e., $\Delta_{a}>0$, there is no quantum entanglement between the auxiliary cavity and the mechanical resonator (the optomechanical cavity), i.e., $E_{a_{a}b}=0$ ($E_{a_{a}a_{c}}=0$). In contrast to this, they become strongly entangled for the blue-detuned case, i.e., $\Delta_{a}<0$, and the highest quantum entanglement can be achieved for $\Delta_{a}/\omega_{m}\approx-1$.

Physically, the combined effect of the optomechanical and tunneling couplings leads to strong entanglement between the indirectly coupled cavity photons and phonons. We also find that the optomechanical entanglement $E_{a_{c}b}$ can be greatly enhanced for the blue-detuned driving of the auxiliary cavity but suppressed for the red-detuned case, and that the maximum entanglement is generated at $\Delta_{a}/\omega_{m}\approx0$.

In particular, the complementary distribution of the entanglement in Figs.~\ref{Entanglement}(b) and~\ref{Entanglement}(a),~\ref{Entanglement}(c) indicates that the initial cooling-cavity-phonon entanglement is partially transferred to the auxiliary-cavity-phonon and photon-photon subsystems. This effect is prominent when the auxiliary-cavity detuning $\Delta_{a}/\omega_{m}=-1$. In additional, Fig.~\ref{Entanglement} shows that the entanglement is higher for a smaller decay rate $\kappa_{a}$ of the auxiliary cavity.

Because the above enhancement of the entanglement results from the ACA mechanism, it is natural to ask the question whether we can further explore the quantum entanglement by tuning the parameters of the ACA mechanism. To further elucidate this aspect, we plot the logarithmic negativities $E_{a_{a}b}$, $E_{a_{c}b}$, and $E_{a_{a}a_{c}}$ as functions of the tunneling coupling $J$ and the pump power $P_{R}$ of the auxiliary cavity, as shown in Figs.~\ref{Entanglement}(d), ~\ref{Entanglement}(e), and ~\ref{Entanglement}(f), respectively. We find that, by using the ACA mechanism, \emph{both photon-phonon and photon-photon entanglement are generated}, and the photon-phonon entanglement is much \emph{larger than} photon-photon entanglement, i.e., $E_{a_{a}b}$, $E_{a_{c}b}>E_{a_{a}a_{c}}$.
The highest entanglement $E_{a_{a}b}$ and $E_{a_{a}a_{c}}$ is observed for: $0.3\leq J/\omega_{m}\leq0.6$ and $45\leq P_{R}$ mW, and $E_{a_{c}b}$ is observed for: $0.15\leq J/\omega_{m}\leq0.3$ and $50\leq P_{R}$ mW. This offers a new method to generate and enhance fragile quantum resources by utilizing auxiliary devices.

\subsection{ACA noise-tolerant ability\label{sec5C}}

Thermal noises in practical devices can destroy fragile quantum resources. To protect quantum resources from environmental thermal perturbations, we introduce the ACA mechanism, which can significantly improve the robustness of quantum entanglement against thermal noises.

\begin{figure}[tbp]
\center
\includegraphics[width=0.45 \textwidth]{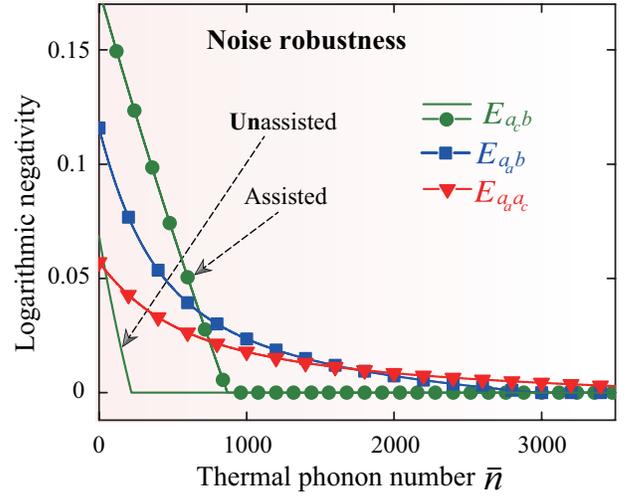}
\caption{Logarithmic negativities $E_{a_{c}b}$ (green), $E_{a_{a}b}$ (blue), and $E_{a_{a}a_{c}}$ (red) versus the thermal phonon numbers $\bar{n}$ of the mechanical resonator in the auxiliary-cavity-unassisted and auxiliary-cavity-assisted cases, when $\kappa_{a}=0.5\kappa_{c}$. Other parameters are the same as those used in Fig.~\ref{Entanglement}.}
\label{thermalnoise}
\end{figure}

When the system works in both auxiliary-cavity-unassisted (the green solid line) and -assisted (marked by green symbols) cases, we plot the logarithmic negativities $E_{a_{a}b}$ and $E_{a_{c}b}$ as a function of the thermal excitation number $\bar{n}$ of the mechanical resonator, as shown in Fig.~\ref{thermalnoise}. We find that the optomechanical entanglement $E_{a_{c}b}$ is greatly improved by the ACA method, and its robustness against thermal noises is much stronger than that of the unassisted case. For example, when we switch the auxiliary-cavity-unassisted to -assisted cases, $E_{a_{c}b}$ can be increased from $E_{a_{c}b}\approx0.07$ to $E_{a_{c}b}\approx0.17$ when $\bar{n}=0$. This means that the ACA mechanism can significantly enhance the optomechanical entanglement.

In addition, we can see from Fig.~\ref{thermalnoise} that, in the unassisted case, quantum entanglement only emerges when $\bar{n}\ll200$ (see the green solid line), while in the assisted case, it can persist for thermal phonons near $\bar{n}=900$ (see the green symbols), which means that the noise robustness in the ACA case is $4.5$ times greater than that in the unassisted case. This indicates that the ACA mechanism provides a feasible method to protect fragile quantum resources from environmental thermal perturbations in practical devices, and to build noise-tolerant quantum processors. Moreover, we find that due to the combined effect of the optomechanical and tunneling couplings, the indirectly coupled cavity photons and phonons can be entangled strongly ($E_{a_{a}b}$, marked by blue symbols), and the robustness of quantum entanglement against noise is even up to three times that of the directly coupled case, as shown in Fig.~\ref{thermalnoise}. In particular, owing to the tunneling coupling between the cooling and auxiliary cavities, the cooling-cavity photons and auxiliary-cavity photons can be entangled, and this entanglement is strongly robust against thermal noise ($E_{a_{a}a_{c}}$, marked by the red symbols). These findings provide a useful strategy to improve the performance of fragile quantum resources by just utilizing auxiliary devices.

\subsection{Tripartite entanglement\label{sec5D}}

\begin{figure}[tbp]
\center
\includegraphics[width=0.45 \textwidth]{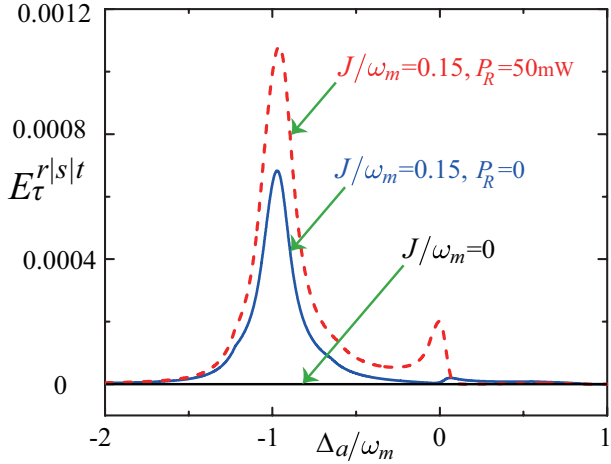}
\caption{The tripartite entanglement, quantified by the minimum residual contangle $E^{r|s|t}_{\tau}$, versus the scaled effective driving detuning $\Delta_{a}/\omega_{m}$ when $J=0$ (the horizontal black solid line), $J/\omega_{m}=0.15$ and $P_{R}=0$ (the blue solid curve), and $J/\omega_{m}=0.15$ and $P_{R}=50$ mW (the red dashed curve), under $\kappa_{a}=0.5\kappa_{c}$. Other parameters are the same as those used in Fig.~\ref{Figmodel}.}
\label{sanchong}
\end{figure}

Besides bipartite entanglements, the application of the ACA mechanism can lead to a genuinely tripartite entanglement, as demonstrated
by the nonzero minimum residual contangle in Eq.~(\ref{minimum}). In Fig.~\ref{sanchong}, we plot the tripartite entanglement, quantified by the minimum residual contangle $E^{r|s|t}_{\tau}$, versus the scaled effective driving detuning $\Delta_{a}/\omega_{m}$ when $J=0$ (see the horizontal black solid line), $J/\omega_{m}=0.15$ and $P_{R}=0$ (blue solid curve), and $J/\omega_{m}=0.15$ and $P_{R}=50$ mW (red dashed curve). We find that, without the ACA mechanism (i.e., $J=0$), no tripartite
entanglement is generated (i.e., $E^{r|s|t}_{\tau}=0$, see the horizontal black solid line); while with the ACA mechanism (i.e., $J/\omega_{m}=0.15$), strong tripartite entanglement is generated (i.e., $E^{r|s|t}_{\tau}>0$, see the blue or red curves). In particular, the tripartite entanglement in $P_{R}\neq0$ case (see the red dashed curve) is much stronger than that when in $P_{R}=0$ case (see the blue solid curve). Very recently, the tripartite entanglement has been achieved in a cavity magnomechanical system, which consists
of cavity microwave photons, magnons, and phonons~\cite{Li2018PRL,Li2019PRA,Yu2020PRL}.

Finally, we remark that in experiments, quantum entanglement can be detected by measuring the covariance matrix $\mathbf{\tilde{V}}$ under a proper readout choice via a filter~\cite{Genes2008PRA1,Riedinger2016Nature,Palomaki2013Science}. The optical quadratures can be measured via the homodyne or heterodyne detection of the output ~\cite{Palomaki2013Science,Barzanjeh2019Nature,Chen2020NC}, and the readout of mechanical quadratures requires a probe being resonant with the anti-Stokes sideband, mapping the mechanical motion to the output field~\cite{Palomaki2013Science}.

\section{Conclusion\label{sec6}}

In conclusion, we have shown how to achieve a giant amplification in the net cooling rate of a mechanical resonator, and to realize significantly enhanced optomechanical refrigeration and entanglement in an auxiliary-cavity-assisted optomechanical system. We have demonstrated that the genuine tripartite entanglement of cooling-cavity photons, auxiliary-cavity photons, and phonons can be generated by using the ACA method. Specifically, we have revealed that the tripartite entanglement arises
from the ACA mechanism, without which it vanishes. We also found that the blue-detuned driving of the auxiliary cavity leads to an enhanced cooling and entanglement, while the red-detuned driving suppresses them.

More importantly, we have revealed that the ACA entanglement has a much stronger robustness against thermal noises in comparison with the auxiliary-cavity-unassisted case. Our work could potentially be used for further manipulating and observing quantum mechanical effects, protecting fragile quantum resources from environmental thermal noises, and building noise-tolerant quantum processors.

\begin{acknowledgments}
B.-P.H. is supported in part by National Natural Science Foundation of China (Grant No.~11974009). A.M. is supported by the Polish National Science Centre (NCN) under the Maestro Grant No. DEC-2019/34/A/ST2/00081. F.N. is supported in part by:
 Nippon Telegraph and Telephone Corporation (NTT) Research,
 the Japan Science and Technology Agency (JST) [via
 the Quantum Leap Flagship Program (Q-LEAP) program,
 the Moonshot R\&D Grant Number JPMJMS2061, and
 the Centers of Research Excellence in Science and Technology (CREST) Grant No. JPMJCR1676],
 the Japan Society for the Promotion of Science (JSPS)
 [via the Grants-in-Aid for Scientific Research (KAKENHI) Grant No. JP20H00134 and the
 JSPS¨CRFBR Grant No. JPJSBP120194828],
 the Army Research Office (ARO) (Grant No. W911NF-18-1-0358),
 the Asian Office of Aerospace Research and Development (AOARD) (via Grant No. FA2386-20-1-4069), and
 the Foundational Questions Institute Fund (FQXi) via Grant No. FQXi-IAF19-06.
\end{acknowledgments}

\appendix

\section{Analytical expressions of the steady-state mean phonon number\label{appendixa}}

In this Appendix, we show the exact analytical expressions of the steady-state average phonon numbers in the mechanical resonator. As shown in Sec.~\ref{sec3C}, by calculating the integral in Eq.~(\ref{specintegral}) for the position and momentum fluctuation spectra, the exact steady-state mean phonon number can be obtained in the form~\cite{Genes2008PRA,Sommer2019PRL}
\begin{equation}
\label{exactcoolresult}
n_{f}=\frac{1}{2}\left( \frac{iD_{6}}{2\Delta_{6}}+\frac{iM_{6}}{2\Delta_{6}}-1\right) .
\end{equation}
Here, we introduce the variables
\begin{widetext}
\begin{eqnarray}
\Delta _{6}
&=&a_{5}\{a_{4}(-a_{1}a_{2}a_{3}+a_{3}^{2}+a_{1}^{2}a_{4})+[-a_{2}a_{3}+a_{1}(a_{2}^{2}-2a_{4})]a_{5}+a_{5}^{2}\}\notag \\
&&-[a_{3}^{3}-a_{1}a_{3}(a_{2}a_{3}+3a_{5})+a_{1}^{2}(a_{3}a_{4}+2a_{2}a_{5})]a_{6}+a_{1}^{3}a_{6}^{2},
\end{eqnarray}
\begin{eqnarray}
D_{6}&=&\big[-a_{3}a_{4}a_{5}+a_{3}^{2}a_{6}+a_{5}(a_{2}a_{5}-a_{1}a_{6})\big]b_{1}+(a_{1}a_{4}a_{5}-a_{5}^{2}-a_{1}a_{3}a_{6})b_{2}\notag \\
&&+(-a_{1}a_{2}a_{5}+a_{3}a_{5}+a_{1}^{2}a_{6})b_{3}+\big[-a_{3}^{2}-a_{1}^{2}a_{4}+a_{1}(a_{2}a_{3}+a_{5})\big]b_{4}  \notag \\
&&+\frac{1}{a_{6}}\big[a_{3}^{2}a_{4}-a_{2}a_{3}a_{5}+a_{5}^{2}+a_{1}^{2}(a_{4}^{2}-a_{2}a_{6})+a_{1}(-a_{2}a_{3}a_{4}+a_{2}^{2}a_{5}-2a_{4}a_{5}+a_{3}a_{6})\big]b_{5},
\end{eqnarray}
and
\begin{eqnarray}
M_{6} &=&\frac{1}{\omega _{m}^{2}}\Bigg\{-\bigg[a_{5}\left(
-a_{2}a_{3}a_{4}+a_{2}^{2}a_{5}+a_{4}(a_{1}a_{4}-a_{0}a_{5})\right)+\left(
-a_{1}a_{3}a_{4}+a_{0}a_{3}a_{5}+a_{2}(a_{3}^{2}-2a_{1}a_{5})\right)
a_{6}+a_{1}^{2}a_{6}^{2}\bigg]b_{1}  \notag \\
&&+[-a_{3}a_{4}a_{5}+a_{3}^{2}a_{6}+a_{5}(a_{2}a_{5}-a_{1}a_{6})]b_{2}+(a_{1}a_{4}a_{5}-a_{5}^{2}-a_{1}a_{3}a_{6})b_{3}\notag \\
&&+(-a_{1}a_{2}a_{5}+a_{3}a_{5}+a_{1}^{2}a_{6})b_{4}+[-a_{3}^{2}-a_{1}^{2}a_{4}+a_{1}(a_{2}a_{3}+a_{5})]b_{5}\Bigg\},
\end{eqnarray}
where the coefficients are defined by:
\begin{eqnarray}
a_{0} &=&1,  \notag \\
a_{1} &=&-i[2(\kappa _{c}+\kappa _{a})+\gamma _{m}],  \notag \\
a_{2} &=&-2J^{2}-2\kappa _{c}(2\kappa _{a}+\gamma _{m})-\kappa _{a}(\kappa
_{a}+2\gamma _{m})-g_{c}-(\kappa _{c}^{2}+\Delta _{a}^{2}),  \notag \\
a_{3} &=&i\{\kappa _{a}^{2}\gamma _{m}+2J^{2}(\kappa _{c}+\kappa _{a}+\gamma
_{m})+\kappa _{c}^{2}(2\kappa _{a}+\gamma _{m})+2\kappa _{a}g_{c}  \notag \\
&&+\gamma _{m}(\Delta ^{2}+\Delta _{a}^{2})+2\kappa _{c}(2\kappa _{a}\gamma
_{m}+\omega _{m}^{2}+f_{a}^{+})\},  \notag \\
a_{4} &=&J^{4}-2|G|^{2}\omega _{m}\Delta +(2\kappa _{a}\gamma _{m}+\omega
_{m}^{2})\Delta ^{2}+f_{a}^{+}g_{c}+2J^{2}[\kappa _{a}\gamma _{m}+\kappa
_{c}(\kappa _{a}+\gamma _{m})+\omega _{m}^{2}-\Delta \Delta _{a}]  \notag \\
&&+\kappa _{c}^{2}(\kappa _{a}^{2}+2\kappa _{a}\gamma _{m}+g_{a})+2\kappa
_{c}(\gamma _{m}f_{a}^{+}+2\kappa _{a}\omega _{m}^{2}),  \notag \\
a_{5} &=&i\{-J^{4}\gamma _{m}+\kappa _{a}[\kappa _{c}\left( -\kappa
_{c}\kappa _{a}\gamma _{m}-2(\kappa _{c}+\kappa _{a})\omega _{m}^{2}\right)
+4|G|^{2}\omega _{m}\Delta -(\kappa _{a}\gamma _{m}+2\omega _{m}^{2})\Delta
^{2}]  \notag \\
&&-(2\kappa _{c}\omega _{m}^{2}+\gamma _{m}f_{c}^{+})\Delta
_{a}^{2}-2J^{2}[\kappa _{a}\omega _{m}^{2}+\kappa _{c}(\kappa _{a}\gamma
_{m}+\omega _{m}^{2})-\gamma _{m}\Delta \Delta _{a}]\},  \notag \\
a_{6} &=&-\omega _{m}\{J^{4}\omega _{m}+2J^{2}(\kappa _{c}\kappa _{a}\omega
_{m}+|G|^{2}\Delta _{a}-\omega _{m}\Delta \Delta _{a})+[-2|G|^{2}\Delta
+\omega _{m}f_{c}^{+}]f_{a}^{+}\},
\end{eqnarray}
and
\begin{eqnarray}
b_{0} &=&0,  \notag \\
b_{1} &=&\gamma _{m}\omega _{m}^{2}(1+2\bar{n}),  \notag \\
b_{2} &=&2\omega _{m}^{2}[|G|^{2}\kappa _{c}-(1+2\bar{n})\gamma
_{m}(2J^{2}-f_{c}^{-}-f_{a}^{-})],  \notag \\
b_{3} &=&\omega _{m}^{2}\{2|G|^{2}[J^{2}(-2\kappa _{c}+\kappa _{a})+\kappa
_{c}(f_{c}^{+}+2f_{a}^{-})]+(1+2\bar{n})\gamma _{m}[6J^{4}-4\Delta
^{2}f_{a}^{-}+f_{c}^{+2}+f_{a}^{+2}  \notag \\
&&+4\kappa _{c}^{2}f_{a}^{-}+4J^{2}(\kappa _{c}\kappa _{a}-\Delta \Delta
_{a}-f_{c}^{-}-f_{a}^{-})]\},  \notag \\
b_{4} &=&2\omega _{m}^{2}\Bigg\{-(1+2\bar{n})\gamma _{m}\bigg\{2J^{6}+\kappa
_{c}^{4}(\Delta _{a}^{2}-\kappa _{a}^{2})+J^{4}(4\kappa _{c}\kappa
_{a}-4\Delta \Delta _{a}-f_{c}^{-}-f_{a}^{-})  \notag \\
&&-2J^{2}[\kappa _{c}\kappa _{a}(\kappa _{c}^{2}-\kappa _{c}\kappa
_{a}+\kappa _{a}^{2})+\kappa _{a}(\kappa _{c}+\kappa _{a})\Delta
^{2}+(4\kappa _{c}\kappa _{a}+\kappa _{a}^{2}+f_{c}^{+})\Delta \Delta _{a}
\notag \\
&&+(\kappa _{c}\kappa _{a}+f_{c}^{-}+\Delta \Delta _{a})\Delta
_{a}^{2}]-\kappa _{c}^{2}(2\Delta ^{2}f_{a}^{-}+f_{a}^{+2})+\Delta
^{2}(f_{a}^{+2}-f_{a}^{-}\Delta ^{2})\bigg\}+|G|^{2}\bigg\{J^{4}(\kappa
_{c}-2\kappa _{a})  \notag \\
&&+\kappa _{c}(2f_{c}^{+}f_{a}^{-}+f_{a}^{+2})+J^{2}\Big[3\kappa _{c}^{2}\kappa
_{a}+\kappa _{a}(\Delta ^{2}+4\Delta \Delta _{a}+f_{a}^{+})+2\kappa (\Delta
\Delta _{a}-f_{a}^{-})\Big]\bigg\}\Bigg\},  \notag \\
b_{5} &=&\omega _{m}^{2}[J^{4}+2J^{2}(\kappa _{c}\kappa _{a}-\Delta \Delta
_{a})+f_{c}^{+}f_{a}^{+}]\Big\{2|G|^{2}(J^{2}\kappa _{a}+\kappa
_{c}f_{a}^{+})+(1+2\bar{n})\gamma _{m}\big[J^{4}+2J^{2}(\kappa _{c}\kappa
_{a}-\Delta \Delta _{a})+f_{c}^{+}f_{a}^{+}\big]\Big\},
\end{eqnarray}%
where%
\begin{eqnarray}
g_{c} =\omega _{m}^{2}+\Delta ^{2}, \hspace{0.5 cm}
g_{a} =\omega _{m}^{2}+\Delta _{a}^{2}, \hspace{0.5 cm}
f_{c}^{\pm } =\kappa _{c}^{2}\pm \Delta ^{2}, \hspace{0.5 cm}
f_{a}^{\pm } =\kappa _{a}^{2}\pm \Delta _{a}^{2}.
\end{eqnarray}%
\end{widetext}

\section{Bistability analysis\label{appendixb}}

By separating the degrees of classical motion from the quantum fluctuations in Eq.~(\ref{Langevineqorig}), the classical-motion equations can be written as
\begin{eqnarray}
\frac{d}{dt}\left\langle a_{c}\right\rangle  &=&-\left[ \kappa _{c}+i\left(
\Delta _{c}-g_{0}\left\langle q\right\rangle \right) \right] \left\langle
a_{c}\right\rangle -i\Omega _{L}-iJ\left\langle a_{a}\right\rangle ,  \notag
\\
\frac{d}{dt}\left\langle a_{a}\right\rangle  &=&-\left( \kappa _{a}+i\Delta
_{a}\right) \left\langle a_{a}\right\rangle -iJ\left\langle
a_{c}\right\rangle -i\Omega _{R},  \notag \\
\frac{d}{dt}\left\langle p\right\rangle  &=&-\omega _{m}\left\langle
q\right\rangle +g_{0}\left\langle a_{c}^{\dagger }\right\rangle \left\langle
a_{c}\right\rangle -\gamma _{m}\left\langle p\right\rangle ,  \notag \\
\frac{d}{dt}\left\langle q\right\rangle  &=&\omega _{m}\left\langle
p\right\rangle.
\end{eqnarray}

\begin{figure}[tbp]
\center
\includegraphics[width=0.45 \textwidth]{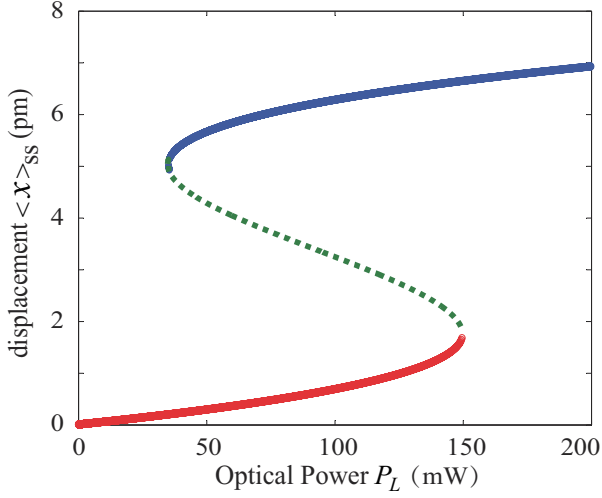}
\caption{The steady-state average displacement $\langle x\rangle _{\text{ss}}$ of the mechanical resonator as a function of the left optical power $P_{L}$. The green dashed curve indicates the unstable solutions. Here we set $\Delta_{c}=\omega_{m}$ and $\Delta_{a}=-\omega_{m}$. Other parameters are the same as those used in Fig.~\ref{Figmodel}.}
\label{bistability}
\end{figure}

The steady-state mean values of the dynamical variables can be obtained as

\begin{eqnarray}
\left\langle a_{c}\right\rangle _{\text{ss}} &=&\frac{i(\Omega
_{L}+J\left\langle a_{a}\right\rangle _{\text{ss}})}{-(\kappa _{c}+i\Delta )}%
,  \notag \\
\left\langle a_{a}\right\rangle _{\text{ss}} &=&\frac{i(\Omega
_{R}+J\left\langle a_{c}\right\rangle _{\text{ss}})}{-(\kappa _{a}+i\Delta
_{a})},  \notag \\
\left\langle p\right\rangle _{\text{ss}} &=&0,  \notag \\
\left\langle q\right\rangle _{\text{ss}} &=&\frac{g_{0}\left\langle
a_{c}^{\dagger }\right\rangle _{\text{ss}}\left\langle a_{c}\right\rangle _{\text{ss}}}{\omega _{m}}, \notag \\
\langle x\rangle _{\text{ss}}&=&\frac{\langle q\rangle _{\text{ss}}}{\sqrt{m\omega_{m}}},\label{steady-sate}
\end{eqnarray}%
where
\begin{eqnarray}
\Delta  &=&\Delta _{c}-g_{0}\left\langle q\right\rangle _{\text{ss}},  \notag
\\
\Delta _{a} &=&\omega _{a}-\omega _{R}.
\end{eqnarray}
In Fig.~\ref{bistability}, we plot the steady-state average displacement $\langle x\rangle _{\text{ss}}$ of the mechanical resonator as a function of the left optical power $P_{L}$. One can see that the steady-state average displacement $\langle x\rangle _{\text{ss}}$ varies with the driving power $P_{L}$ of the left driving field by solving Eqs.~(\ref{steady-sate}) numerically. It is shown that when $P_{L}<35$ mW, only one solution of $\langle x\rangle _{\text{ss}}$ exists and the system is not bistable. When $35<P_{L}<150$ mW, three solutions of $\langle x\rangle _{\text{ss}}$ exist and the green dashed curve corresponds to the unstable solutions. So the system exhibits bistability in this case. To obtain the cooling and entanglement, a single solution region should be chosen, and we set $P_{L}<35$ mW throughout this work.


\begin{thebibliography}{99}

\bibitem{Kippenberg2008Science}       T. J. Kippenberg and K. J. Vahala, Cavity Optomechanics: Back-Action at the Mesoscale, Science \textbf{321}, 1172 (2008).

\bibitem{Meystre2013AP}       P. Meystre, A short walk through quantum optomechanics, Ann. Phys. (Berlin) \textbf{525}, 215 (2013).

\bibitem{Aspelmeyer2014RMP}       M. Aspelmeyer, T. J. Kippenberg, and F. Marquardt, Cavity optomechanics, Rev. Mod. Phys. \textbf{86}, 1391  (2014).

\bibitem{Bowen2015book}  W. P. Bowen and G. J. Milburn, Quantum Optomechanics, (Boca Raton, FL: CRC Press) (2015).


\bibitem{Rabl2011PRL}       P. Rabl, Photon Blockade Effect in Optomechanical Systems, Phys. Rev. Lett. \textbf{107}, 063601 (2011).

\bibitem{Nunnenkamp2011}   A. Nunnenkamp, K. B{\o}rkje, and S. M. Girvin, Single-Photon Optomechanics, Phys. Rev. Lett. \textbf{107}, 063602 (2011).

\bibitem{Liao2012PRA}         J.-Q. Liao, H. K. Cheung, and C. K. Law, Spectrum of single-photon emission and scattering in cavity optomechanics, Phys. Rev. A \textbf{85}, 025803 (2012).

\bibitem{Liao2013PRA}       J.-Q. Liao and F. Nori, Photon blockade in quadratically coupled optomechanical systems, Phys. Rev. A \textbf{88}, 023853 (2013).


\bibitem{Wang2015PRA1}       H. Wang, X. Gu, Y.-x. Liu, A. Miranowicz, and F. Nori, Tunable photon blockade in a hybrid system consisting of an optomechanical device coupled to a two-level system, Phys. Rev. A \textbf{92}, 033806 (2015).


\bibitem{Huang2018PRL}  R. Huang, A. Miranowicz, J.-Q. Liao, F. Nori, and H. Jing, Nonreciprocal Photon Blockade, Phys. Rev. Lett. \textbf{121}, 153601 (2018).

\bibitem{Li2019PR}  B. Li, R. Huang, X. Xu, A. Miranowicz, and H. Jing, Nonreciprocal unconventional photon blockade in a spinning optomechanical system, Photonics Res. \textbf{7}, 630 (2019).

\bibitem{Zou2019PRA}       F. Zou, L.-B. Fan, J.-F. Huang, and J.-Q. Liao, Enhancement of few-photon optomechanical effects with cross-Kerr nonlinearity, Phys. Rev. A \textbf{99}, 043837 (2019).

\bibitem{Liao2020PRA}       J.-Q. Liao, J.-F. Huang, L. Tian, L.-Ma. Kuang, and C.-P. Sun, Generalized ultrastrong optomechanical-like coupling, Phys. Rev. A \textbf{101}, 063802 (2020).



\bibitem{Agarwal2010PRA}       G. S. Agarwal and S. Huang, Electromagnetically induced transparency in mechanical effects of light, Phys. Rev. A \textbf{81}, 041803(R) (2010).

\bibitem{Weis2010Science}       S. Weis, R. Rivi\`{e}re, S. Del\'{e}glise, E. Gavartin, O. Arcizet, A. Schliesser, and T. J. Kippenberg, Optomechanically induced transparency, Science \textbf{330}, 1520 (2010).

\bibitem{Safavi-Naeini2011Nature}       A. H. Safavi-Naeini, T. P. M. Alegre, J. Chan, M. Eichenfield, M. Winger, Q. Lin, J. T. Hill, D. E. Chang, and O. Painter, Electromagnetically induced transparency and slow light with optomechanics, Nature (London) \textbf{472}, 69 (2011).

    \bibitem{Wang2014PRA}       H. Wang, X. Gu, Y.-x. Liu, A. Miranowicz, and F. Nori, Optomechanical analog of two-color electromagnetically induced transparency: Photon transmission through an optomechanical device with a two-level system, Phys. Rev. A \textbf{90}, 023817  (2014).

\bibitem{Hou2015PRA}       B. P. Hou, L. F. Wei, and S. J. Wang, Optomechanically induced transparency and absorption in hybridized optomechanical systems, Phys. Rev. A \textbf{92}, 033829  (2015).

\bibitem{Lai2020PRA1}       D.-G. Lai, X. Wang, W. Qin, B.-P. Hou, F. Nori, and J.-Q. Liao, Tunable optomechanically induced transparency by controlling the dark-mode effect, Phys. Rev. A  \textbf{102}, 023707 (2020).



\bibitem{Cirio2017PRL} M. Cirio, K. Debnath, N. Lambert, and F. Nori, Amplified Optomechanical Transduction of Virtual Radiation Pressure, Phys. Rev. Lett. \textbf{119}, 053601 (2017).

 \bibitem{Stefano2019PRL}     O. D. Stefano, A. Settineri, V. Macr\`{\i}, A. Ridolfo, R. Stassi, A. F. Kockum, S. Savasta, and F. Nori, Interaction of Mechanical Oscillators Mediated by the Exchange of Virtual Photon Pairs, Phys. Rev. Lett. \textbf{122}, 030402 (2019).


\bibitem{Qin2019PRA}       W. Qin, V. Macr\`{\i}, A. Miranowicz, S. Savasta, and F. Nori, Emission of photon pairs by mechanical stimulation of the squeezed vacuum, Phys. Rev. A \textbf{100}, 062501 (2019).

\bibitem{Wang2019PRA}       X. Wang, W. Qin, A. Miranowicz, S. Savasta, and F. Nori, Unconventional cavity optomechanics: Nonlinear control of phonons in the acoustic quantum vacuum, Phys. Rev. A \textbf{100}, 063827 (2019).



     \bibitem{Malz2018PRL} D. Malz, L. D. T\'{o}th, N. R. Bernier, A. K. Feofanov, T. J. Kippenberg, and A. Nunnenkamp, Quantum-Limited Directional Amplifiers with Optomechanics, Phys. Rev. Lett. \textbf{120}, 023601 (2018).

\bibitem{Shen2016NP} Z. Shen, Y.-L. Zhang, Y. Chen, C.-L. Zou, Y.-F. Xiao, X.-B. Zou, F.-W. Sun, G.-C. Guo, and C.-H. Dong, Experimental realization of optomechanically induced non-reciprocity, Nat. Photonics \textbf{10}, 657 (2016).

\bibitem{Shen2018NC} Z. Shen, Y.-L. Zhang, Y. Chen, F.-W. Sun, X.-B. Zou, G.-C. Guo, C.-L. Zou, and C.-H. Dong, Reconfigurable optomechanical circulator and directional amplifier, Nat. Commun. \textbf{9}, 1797 (2018).


   \bibitem{Fang2017NP} K. Fang, J. Luo, A. Metelmann, M. H. Matheny, F. Marquardt, A. A. Clerk, and O. Painter, Generalized non-reciprocity in an optomechanical circuit via synthetic magnetism and reservoir engineering, Nat. Phys. \textbf{13}, 465 (2017).

\bibitem{Xu2019Nature} H. Xu, L. Jiang, A. A. Clerk, and J. G. E. Harris, Nonreciprocal control and cooling of phonon modes in an optomechanical system, Nature (London) \textbf{568}, 65 (2019).

\bibitem{Mathew2018arXiv}       J. P. Mathew, J. d. Pino, and E. Verhagen, Synthetic gauge fields for phonon transport in a nano-optomechanical system, Nat. Nanotechnol. \textbf{15}, 198 (2020).

    \bibitem{Yang2020NC}    C. Yang, X. Wei, J. Sheng, and H. Wu, Phonon heat transport in cavity-mediated optomechanical nanoresonators, Nat. Commun. \textbf{11}, 4656 (2020).

    \bibitem{Xu2016Nature} H. Xu, D. Mason, L. Jiang, and J. G. E. Harris, Topological energy transfer in an optomechanical system with exceptional points, Nature (London) \textbf{537}, 80 (2016).

        \bibitem{SanavioPRB2020} C. Sanavio, V. Peano, and A. Xuereb, Nonreciprocal topological phononics in optomechanical arrays, Phys. Rev. B \textbf{101}, 085108 (2020).




\bibitem{Wilson-Rae2007PRL}       I. Wilson-Rae, N. Nooshi, W. Zwerger, and T. J. Kippenberg, Theory of Ground State Cooling of a Mechanical Oscillator Using Dynamical Backaction, Phys. Rev. Lett. \textbf{99}, 093901 (2007).

\bibitem{Marquardt2007PRL}       F. Marquardt, J. P. Chen, A. A. Clerk, and S. M. Girvin, Quantum Theory of Cavity-Assisted Sideband Cooling of Mechanical Motion, Phys. Rev. Lett. \textbf{99}, 093902 (2007).

\bibitem{Genes2008PRA}      C. Genes, D. Vitali, P. Tombesi, S. Gigan, and M. Aspelmeyer, Ground-state cooling of a micromechanical oscillator: Comparing cold damping and cavity-assisted cooling schemes, Phys. Rev. A \textbf{77}, 033804 (2008).


\bibitem{Vitali2007PRL}       D. Vitali, S. Gigan, A. Ferreira, H. R. B\"{o}hm, P. Tombesi, A. Guerreiro, V. Vedral, A. Zeilinger, and M. Aspelmeyer, Optomechanical Entanglement between a Movable Mirror and a Cavity Field, Phys. Rev. Lett. \textbf{98}, 030405 (2007).


\bibitem{Genes2008NJP} C. Genes, D. Vitali, and P. Tombesi, Simultaneous cooling and entanglement of mechanical modes of a micromirror in an optical cavity, New J. Phys. \textbf{10}, 095009 (2008).


\bibitem{Vitali2007JPA}      D. Vitali, S. Mancini and P. Tombesi, Stationary entanglement between two movable mirrors in a classically driven Fabry-Perot cavity, J. Phys. A: Math. Theor. \textbf{40}, 8055 (2007).


\bibitem{Paternostro2007PRL} M. Paternostro, D. Vitali, S. Gigan, M. S. Kim, C. Brukner, J. Eisert, and M. Aspelmeyer, Creating and Probing Multipartite Macroscopic Entanglement with Light, Phys. Rev. Lett. \textbf{99}, 250401 (2007).

\bibitem{Mancini2002PRL}      S. Mancini, V. Giovannetti, D. Vitali, and P. Tombesi, Entangling Macroscopic Oscillators Exploiting Radiation Pressure, Phys. Rev. Lett. \textbf{88}, 120401 (2002).


\bibitem{Riedinger2018Nature}  R. Riedinger, A. Wallucks, I. Marinkovi\'{c}, C. L\"{o}schnauer, M. Aspelmeyer, S. Hong, and S. Gr\"{o}blacher, Remote quantum entanglement between two micromechanical oscillators, Nature (London) \textbf{556}, 473 (2018).

\bibitem{Ockeloen-Korppi2018Nature}       C. F. Ockeloen-Korppi, E. Damsk\"{a}gg, J.-M. Pirkkalainen,  M. Asjad, A. A. Clerk, F. Massel, M. J. Woolley, and M. A. Sillanp\"{a}\"{a}, Stabilized entanglement of massive mechanical oscillators, Nature (London) \textbf{556}, 478 (2018).

\bibitem{Qin2019npj}       W. Qin, A. Miranowicz, G. L. Long, J. Q. You, and F. Nori, Proposal to test quantum wave-particle superposition on massive mechanical resonators, npj Quantum Information \textbf{5}, 58 (2019).




\bibitem{Mancini1998PRL} S. Mancini, D. Vitali, and P. Tombesi, Optomechanical Cooling of a Macroscopic Oscillator by Homodyne Feedback, Phys. Rev. Lett. \textbf{80}, 688 (1998).


\bibitem{Steixner2005PRA} V. Steixner, P. Rabl, and P. Zoller, Quantum feedback cooling of a single trapped ion in front of a mirror, Phys. Rev. A \textbf{72}, 043826 (2005).

\bibitem{Bushev2006PRL} P. Bushev, D. Rotter, A. Wilson, F. M. C. Dubin, C. Becher, J. Eschner, R. Blatt, V. Steixner, P. Rabl, and P. Zoller, Feedback Cooling of a Single Trapped Ion, Phys. Rev. Lett. \textbf{96}, 043003 (2006).


    \bibitem{Rossi2017PRL} M. Rossi, N. Kralj, S. Zippilli, R. Natali, A. Borrielli, G. Pandraud, E. Serra, G. D. Giuseppe, and D. Vitali, Enhancing Sideband Cooling by Feedback-Controlled Light, Phys. Rev. Lett. \textbf{119}, 123603 (2017).

\bibitem{Rossi2018Nature} M. Rossi, D. Mason, J. Chen, Y. Tsaturyan, and A. Schliesser, Measurement-based quantum control of mechanical motion, Nature (London) \textbf{563}, 53 (2018).
\bibitem{Conangla2019PRL} G. P. Conangla, F. Ricci, M. T. Cuairan, A. W. Schell, N. Meyer, and R. Quidant, Optimal Feedback Cooling of a Charged Levitated Nanoparticle with Adaptive Control, Phys. Rev. Lett. \textbf{122}, 223602 (2019).

\bibitem{Tebbenjohanns2019PRL} F. Tebbenjohanns, M. Frimmer, A. Militaru, V. Jain, and L. Novotny, Cold Damping of an Optically Levitated Nanoparticle to Microkelvin Temperatures, Phys. Rev. Lett. \textbf{122}, 223601 (2019).

\bibitem{Sommer2019PRL} C. Sommer and C. Genes, Partial Optomechanical Refrigeration via Multimode Cold-Damping Feedback, Phys. Rev. Lett. \textbf{123}, 203605 (2019).

\bibitem{Guo2019PRL} J. Guo, R. Norte, and S. Gr\"{o}blacher, Feedback Cooling of a Room Temperature Mechanical Oscillator close to its Motional Ground State, Phys. Rev. Lett. \textbf{123}, 223602 (2019).

    \bibitem{Sommer2020PRR}       C. Sommer, A. Ghosh, and C. Genes, Multimode cold-damping optomechanics with delayed feedback, Phys. Rev. Research \textbf{2} 033299 (2020).




\bibitem{Wang2011PRL}       X.-T. Wang, S. Vinjanampathy, F. W. Strauch, and K. Jacobs, Ultraefficient Cooling of Resonators: Beating Sideband Cooling with Quantum Control, Phys. Rev. Lett. \textbf{107}, 177204 (2011).

\bibitem{Li2011PRB}       Y. Li, L.-A. Wu, Y.-D. Wang, and L.-P. Yang, Nondeterministic ultrafast ground-state cooling of a mechanical resonator, Phys. Rev. B \textbf{84}, 094502 (2011).

\bibitem{Yan2016PRA}       L.-L. Yan, J.-Q. Zhang, S. Zhang, and M. Feng, Efficient cooling of quantized vibrations using a four-level configuration, Phys. Rev. A \textbf{94}, 063419 (2016).


\bibitem{Liuyu2017PRA}       Y.-L. Liu and Y.-X. Liu, Energy-localization-enhanced groundstate cooling of a mechanical resonator from room temperature
in optomechanics using a gain cavity, Phys. Rev. A \textbf{96}, 023812 (2017).

\bibitem{Liao2011PRA}       J.-Q. Liao and C. K. Law, Cooling of a mirror in cavity optomechanics with a chirped pulse, Phys. Rev. A \textbf{84}, 053838 (2011).

\bibitem{Machnes2012PRL}       S. Machnes, J. Cerrillo, M. Aspelmeyer, W. Wieczorek, M. B. Plenio, and A. Retzker, Pulsed Laser Cooling for Cavity Optomechanical Resonators, Phys. Rev. Lett. \textbf{108}, 153601 (2012).

\bibitem{Lai2018PRA} D.-G. Lai, F. Zou, B.-P. Hou, Y.-F. Xiao, and J.-Q. Liao, Simultaneous cooling of coupled mechanical resonators in cavity optomechanics, Phys. Rev. A \textbf{98}, 023860 (2018).

    \bibitem{Lai2021PRA} D.-G. Lai, J. Huang, B.-P. Hou, F. Nori, and J.-Q. Liao, Domino cooling of a coupled mechanical-resonator chain via cold-damping feedback, Phys. Rev. A \textbf{103}, 063509 (2021).


\bibitem{Liu2013PRL}       Y.-C. Liu, Y.-F. Xiao, X. Luan, and C. W. Wong, Dynamic Dissipative Cooling of a Mechanical Resonator in Strong Coupling Optomechanics, Phys. Rev. Lett. \textbf{110}, 153606 (2013)

\bibitem{Liu2014PRA}       Y.-C. Liu, Y.-F. Shen, Q. Gong, and Y.-F. Xiao, Optimal limits of cavity optomechanical cooling in the strong-coupling regime, Phys. Rev. A \textbf{89}, 053821 (2014).

\bibitem{Lai2020PRARC}       D.-G. Lai, J.-F. Huang, X.-L. Yin, B.-P. Hou, W. Li, D. Vitali, F. Nori, and J.-Q. Liao, Nonreciprocal ground-state cooling of multiple mechanical resonators, Phys. Rev. A  \textbf{102}, 011502(R) (2020).





    \bibitem{Chan2011Nature}       J. Chan, T. P. Alegre, A. H. Safavi-Naeini, J. T. Hill, A. Krause, S. Groeblacher, M. Aspelmeyer, and O. Painter, Laser cooling of a nanomechanical oscillator into its quantum ground state, Nature (London) \textbf{478}, 89 (2011).

\bibitem{Teufel2011Nature}    J. D. Teufel, T. Donner, D. Li, J. W. Harlow, M. S. Allman, K. Cicak, A. J. Sirois, J. D. Whittaker, K. W. Lehnert, and R. W. Simmonds, Sideband cooling of micromechanical motion to the quantum ground state, Nature (London) \textbf{475}, 359 (2011).

\bibitem{Clarkl2017Nature}   J. B. Clark, F. Lecocq, R. W. Simmonds, J. Aumentado, and J. D. Teufel, Sideband cooling beyond the quantum backaction limit with squeezed light, Nature (London) \textbf{541}, 191 (2017).

 \bibitem{MXu2020PRL}   M. Xu, X. Han, C.-L. Zou, W. Fu, Y. Xu, C. Zhong, L. Jiang, and H. X. Tang, Radiative Cooling of a Superconducting Resonator, Phys. Rev. Lett. \textbf{124}, 033602 (2020).

 \bibitem{Qiu2020PRL}   L. Qiu, I. Shomroni, P. Seidler, and T. J. Kippenberg, Laser Cooling of a Nanomechanical Oscillator to Its Zero-Point Energy, Phys. Rev. Lett. \textbf{124}, 173601 (2020).





\bibitem{Xue2007PRB}       F. Xue, Y. D. Wang, Y. X. Liu, and F. Nori, Cooling a micro-mechanical beam by coupling it to a transmission line, Phys. Rev. B \textbf{76}, 205302 (2007).

  \bibitem{You2008PRL} J. Q. You, Y. X. Liu, and F. Nori, Simultaneous cooling of an artificial atom and its neighboring quantum system, Phys. Rev. Lett. \textbf{100}, 047001 (2008).

\bibitem{Zhang2009PRA} J. Zhang, Y. X. Liu, and F. Nori, Cooling and squeezing the fluctuations of a nanomechanical beam by indirect quantum feedback control, Phys. Rev. A \textbf{79}, 052102 (2009).

    \bibitem{Liberato2011PRA} S. D. Liberato, N. Lambert, and F. Nori, Quantum noise in photothermal cooling, Phys. Rev. A \textbf{83}, 033809 (2011).

\bibitem{Grajcar2008PRB}       M. Grajcar, S. Ashhab, J. R. Johansson, and F. Nori, Lower limit on the achievable temperature in resonator-based sideband cooling, Phys. Rev. B \textbf{78}, 035406 (2008).

\bibitem{Nori2008NP}  F. Nori, Atomic physics with a circuit, Nat. Physics \textbf{4}, 589 (2008).

\bibitem{Xiang2013RMP} Z. L. Xiang, S. Ashhab, J. Q. You, and F. Nori, ``Hybrid quantum circuits: Superconducting circuits interacting with other quantum systems", Rev. Mod. Phys. \textbf{85}, 623 (2013).





\bibitem{Hofer2011PRA} S. G. Hofer, W. Wieczorek, M. Aspelmeyer, and K. Hammerer, Phys. Rev. A \textbf{84}, 052327 (2011).

\bibitem{Vanner2011PNAS} M. R. Vanner, I. Pikovski, G. D. Cole, M. S. Kim, C. Brukner, K. Hammerer, G. J. Milburn, and M. Aspelmeyer, Proc. Natl. Acad. Sci. U.S.A. \textbf{108}, 16182 (2011).


\bibitem{Wang2013PRL}   Y.-D. Wang and A. A. Clerk, Reservoir-Engineered Entanglement in Optomechanical Systems, Phys. Rev. Lett. \textbf{110}, 253601 (2013).

\bibitem{Chen2014PRA}  R.-X. Chen, L.-T. Shen, Z.-B. Yang, H.-Z. Wu, and S.-B. Zheng, Phys. Rev. A \textbf{89}, 023843 (2014).

\bibitem{Wang2015PRA} Y.-D. Wang, S. Chesi, and A. A. Clerk, Bipartite and tripartite output entanglement in three-mode optomechanical systems, Phys. Rev. A \textbf{91}, 013807 (2015).

\bibitem{Woolley2014PRA} M. J. Woolley and A. A. Clerk, Two-mode squeezed states in cavity optomechanics via engineering of a single reservoir, Phys. Rev. A \textbf{89}, 063805 (2014).

\bibitem{Yang2015PRA} C.-J. Yang, J.-H. An, W. Yang, and Y. Li, Generation of stable entanglement between two cavity mirrors by squeezed-reservoir engineering, Phys. Rev. A \textbf{92}, 062311 (2015).

\bibitem{Li2015NJP} J. Li, I. M. Haghighi, N. Malossi, S. Zippilli, and D. Vitali, Generation and detection of large and robust entanglement between two different mechanical resonators in cavity optomechanics, New J. Phys. \textbf{17}, 103037 (2015).

\bibitem{LiaoCG2018PRA} C.-G. Liao, R.-X. Chen, H. Xie, and X.-M. Lin, Reservoir-engineered entanglement in a hybrid modulated three-mode optomechanical system, Phys. Rev. A \textbf{97}, 042314 (2018).




\bibitem{Tian2013PRL}   L. Tian, Robust Photon Entanglement via Quantum Interference in Optomechanical Interfaces, Phys. Rev. Lett. \textbf{110}, 233602 (2013).

\bibitem{Genes2011PRA}       C. Genes, H. Ritsch, M. Drewsen, and A. Dantan, Atom-membrane cooling and entanglement using cavity electromagnetically induced transparency, Phys. Rev. A \textbf{84}, 051801(R) (2011).

    \bibitem{Guo2014PRA}   Y. Guo, K. Li, W. Nie, and Y. Li, Electromagnetically-induced-transparency-like ground-state cooling in a double-cavity optomechanical system, Phys. Rev. A \textbf{90}, 053841 (2014).

      \bibitem{Liu2015PRA}  Y. C. Liu, Y. F. Xiao, X. Luan, Q. Gong, and C. W. Wong, Coupled cavities for motional ground-state cooling and strong optomechanical coupling, Phys. Rev. A \textbf{91}, 033818 (2015).

\bibitem{Gu2013PRA} W. J. Gu and G. X. Li, Quantum interference effects on ground-state optomechanical cooling, Phys. Rev. A \textbf{87}, 025804 (2013).

\bibitem{Feng2017PRA} J.-S. Feng, L. Tan, H.-Q. Gu, and W.-M. Liu, Auxiliary-cavity-assisted ground-state cooling of an optically levitated nanosphere in the
unresolved-sideband regime, Phys. Rev. A \textbf{96}, 063818 (2017).


  \bibitem{Mari2009PRL}   A. Mari and J. Eisert, Gently Modulating Optomechanical Systems, Phys. Rev. Lett. \textbf{103}, 213603 (2009).

 \bibitem{Mari2012NJP}  A. Mari and J. Eisert, Opto- and electro-mechanical entanglement improved by modulation, New J. Phys. \textbf{14}, 075014 (2012).


\bibitem{ZLi2015PRA} Z. Li, S.-l. Ma, and F.-l. Li, Generation of broadband two-mode squeezed light in cascaded double-cavity optomechanical systems, Phys. Rev. A \textbf{92}, 023856 (2015).

\bibitem{Wangm2016PRA} M. Wang, X.-Y. L\"{u}, Y.-D. Wang, J. Q. You, and Y. Wu, Macroscopic quantum entanglement in modulated optomechanics, Phys. Rev. A \textbf{94}, 053807 (2016).



 \bibitem{Ho2018PRL}       M. Ho, E. Oudot, J.-D. Bancal, and N. Sangouard, Witnessing Optomechanical Entanglement with Photon Counting, Phys. Rev. Lett. \textbf{121}, 023602 (2018).

 \bibitem{Jiao2020PRL}       Y.-F. Jiao, S.-D. Zhang, Y.-L. Zhang, A. Miranowicz, L.-M. Kuang, and H. Jing, Nonreciprocal Optomechanical Entanglement against Backscattering Losses, Phys. Rev. Lett. \textbf{125}, 143605 (2020).
\bibitem{Chen2021PRL} Y. Chen, Y.-L. Zhang, Z. Shen, C.-L. Zou, G.-C. Guo, and C.-H. Dong, Synthetic Gauge Fields in a Single Optomechanical Resonator, Phys. Rev. Lett. \textbf{126}, 123603 (2021).

 \bibitem{Landau1958NY}   L. Landau and E. Lifshitz, \emph{Statistical Physics} (Pergamon, New York, 1958).



\bibitem{Kim2019OP} S. Kim, J. M. Taylor, and G. Bahl, Dynamic suppression of Rayleigh backscattering in dielectric resonators, Optica \textbf{6}, 1016 (2019).

\bibitem{Li2008PRB} Y. Li, Y.-D. Wang, F. Xue, and C. Bruder, Quantum theory of transmission line resonator-assisted cooling of a micromechanical resonator, Phys. Rev. B \textbf{78}, 134301 (2008).

    \bibitem{Vidal2002PRA} G. Vidal and R. F. Werner, Computable measure of entanglement, Phys. Rev. A \textbf{65}, 032314 (2002).

\bibitem{Plenio2005PRL} M. B. Plenio, Logarithmic Negativity: A Full Entanglement Monotone That is not Convex, Phys. Rev. Lett. \textbf{95}, 090503 (2005).

\bibitem{Adesso2004PRA} G. Adesso, A. Serafini, and F. Illuminati, Extremal entanglement and mixedness in continuous variable systems, Phys. Rev. A \textbf{70}, 022318 (2004).
\bibitem{Vidal2002PRA} G. Vidal and R. F. Werner, Computable measure of entanglement, Phys. Rev. A \textbf{65}, 032314 (2002); M. B. Plenio, Logarithmic Negativity: A Full Entanglement Monotone That is not Convex, Phys. Rev. Lett. \textbf{95}, 090503 (2005).

\bibitem{Adesso2007JPA} G. Adesso and F. Illuminati, Entanglement in continuous-variable systems: recent advances and current perspectives, J. Phys. A \textbf{40}, 7821 (2007); Continuous variable tangle, monogamy inequality, and entanglement sharing in Gaussian states of continuous variable systems, New J. Phys. \textbf{8}, 15 (2006); Entanglement sharing: from qubits to Gaussian states, Int. J. Quantum Inf., \textbf{3} 383 (2006).

\bibitem{Coffman2000PRA} V. Coffman, J. Kundu, and W. K. Wootters, Distributed entanglement, Phys. Rev. A \textbf{61}, 052306 (2000).

\bibitem{Li2018PRL} J. Li, S.-Y. Zhu, and G. S. Agarwal, Magnon-Photon-Phonon Entanglement in Cavity Magnomechanics, Phys. Rev. Lett. \textbf{121}, 203601 (2018).

\bibitem{Li2019PRA} J. Li, S.-Y. Zhu, and G. S. Agarwal, Squeezed states of magnons and phonons in cavity magnomechanic, Phys. Rev. A \textbf{99}, 021801(R) (2019).

\bibitem{Yu2020PRL} M. Yu, H. Shen, and J. Li, Magnetostrictively Induced Stationary Entanglement between Two Microwave Fields, Phys. Rev. Lett. \textbf{124}, 213604 (2020).



\bibitem{Genes2008PRA1} C. Genes, A. Mari, P. Tombesi, and D. Vitali, Robust entanglement of a micromechanical resonator with output optical fields, Phys. Rev. A \textbf{78}, 032316 (2008).

\bibitem{Riedinger2016Nature} R. Riedinger, S. Hong, R. A. Norte, J. A. Slater, J. Shang, A. G. Krause, V. Anant, M. Aspelmeyer, and S. Gr\"{o}blacher, Non-classical correlations between single photons and phonons from a mechanical oscillator, Nature (London) \textbf{530}, 313 (2016).


\bibitem{Palomaki2013Science} T. A. Palomaki, J. D. Teufel, R. W. Simmonds, and K. W. Lehnert, Entangling mechanical motion with microwave fields, Science \textbf{342}, 710 (2013).


\bibitem{Barzanjeh2019Nature} S. Barzanjeh, E. S. Redchenko, M. Peruzzo, M. Wulf, D. P. Lewis, G. Arnold, and J. M. Fink, Stationary entangled radiation from micromechanical motion, Nature (London) \textbf{570}, 480 (2019).

\bibitem{Chen2020NC} J. Chen, M. Rossi, D. Mason, and A. Schliesser, Entanglement of propagating optical modes via a mechanical interface, Nat. Commun. \textbf{11}, 943 (2020).


\end{thebibliography}
\end{document}